\documentclass[aps,prd,onecolumn,groupedaddress,nofootinbib,preprintnumbers,superscriptaddress]{revtex4-2}  
\usepackage{graphicx,mathtools}
\usepackage{array}
\usepackage{tabularx}
\usepackage{natbib}
\usepackage{epstopdf}
\usepackage{amsmath}
\usepackage{amsfonts}
\usepackage[colorlinks=true,citecolor=red,linkcolor=blue,breaklinks=true]{hyperref}
\usepackage{accents}
\usepackage[figure, table]{hypcap}
\usepackage{amssymb}
\usepackage{appendix}
\usepackage{comment}
\usepackage{bbold}
\usepackage{xcolor}
\usepackage{slashed}
\usepackage{subfigure}
\usepackage{setspace}
\usepackage{footnote}
\usepackage{multirow}
\usepackage{braket}
\usepackage[normalem]{ulem}
\usepackage[utf8]{inputenc}
\usepackage{mathrsfs}
\usepackage{longtable}
\usepackage{enumitem}
\usepackage{orcidlink}
\usepackage{booktabs} 
\usepackage{subfigure}
\usepackage{xcolor}

\allowdisplaybreaks

\definecolor{palatd}{RGB}{104, 36, 109}
\definecolor{palatb}{RGB}{0, 56, 168}
\definecolor{palatr}{rgb}{0.745,0.118,0.176}
\newcommand\myshade{80}
\colorlet{mylinkcolor}{palatr}
\colorlet{mycitecolor}{palatb}
\colorlet{myurlcolor}{palatd}

\hypersetup{
  linkcolor  = mylinkcolor!\myshade!black,
  citecolor  = mycitecolor!\myshade!black,
  urlcolor   = myurlcolor!\myshade!black,
  colorlinks = true
}

\newcommand{\MS}[1]{\textcolor{black}{#1}}

\begin{document}

\title{Dark Matter Capture in Supernovae Modifies Dark Photon Cooling Bounds}

\author{Aritra Gupta\,\orcidlink{0000-0003-2265-9730}}
\email{aritra.gupta@krea.edu.in}
\affiliation{Division of Sciences, Krea University, 5655 Central Expressway, Sri City 517646, India.}

\author{Manibrata Sen\,\orcidlink{0000-0001-7948-4332}}
\email{manibrata@iitb.ac.in}
\affiliation{Indian Institute of Technology Bombay, Bombay Powai, Mumbai 400076, India\\}

\begin{abstract}
Core-collapse supernovae serve as powerful probes of light, weakly coupled particles, such as dark photons. The conventional SN1987A cooling bound constrains the dark photon mass-mixing parameter space by requiring that the luminosity from the proto-neutron star core not exceed the observed neutrino emission. In this work, we revisit these limits by including the effect of dark matter (DM) captured inside the progenitor star before collapse. The trapped DM acts as an additional scattering target for dark photons, modifying their free-streaming length and, consequently, the supernova cooling rate. We perform a self-consistent analysis for both annihilating and asymmetric DM scenarios, incorporating light-mediator effects in the capture rate calculation. For annihilating DM, the equilibrium density remains too small to affect the bounds significantly. In contrast, asymmetric DM can accumulate to large densities, leading to the formation of a ``dark photosphere’’ that suppresses the dark-photon luminosity and reopens regions of parameter space previously excluded by supernova cooling. The results presented are intended as a proof-of-principle demonstration of how astrophysical dark matter populations can alter supernova cooling constraints and do not provide precision exclusion limits on any given model.
\end{abstract}

\maketitle
\section{Introduction}
\label{sec:1}

A core-collapse supernova (SN) remains one of the most extreme astrophysical laboratories to test the tenets of the Standard Model (SM). The meager number of neutrino events observed during SN1987A - a blue supergiant going SN in the Large Magellanic Cloud in February 1987~\cite{Kamiokande-II:1987idp, Bionta:1987qt,1987ESOC...26..237A}, already confirms our basic understanding of the working machinery of a SN. Hence, the data can be used to impose stringent constraints on physics beyond the SM~\cite{Horiuchi:2018ofe,Volpe:2023met,Sen:2024fxa}. This is particularly relevant for testing the presence of ``feebly'' interacting particles, which are otherwise difficult to detect in terrestrial laboratories~\cite{Raffelt:1987yt}.

One of the most significant constraints derived from SN1987A observations is the anomalous cooling bound~\cite{Ellis:1987pk,Raffelt:1987yt,PhysRevLett.60.1797, Burrows:1990pk,Raffelt:1996wa,2025JCAP...01..061F}. The neutrino data from SN1987A provide a way to limit the energy that could be carried away by any new weakly interacting particles. If these hypothetical particles are produced within the supernova core, they could serve as additional channels for energy loss, enhancing the cooling rate. Based on this idea, a threshold for the luminosity of such particles was established. If the luminosity of a new particle species exceeds $\mathcal{L} \gtrsim 3 \times 10^{52} \,{\rm erg\,s}^{-1}$ for a core density of $\rho = 3 \times 10^{14} \,{\rm g/cm^3}$ and temperature $T = 30 \,{\rm MeV}$, the duration of the neutrino burst would be shorter than the observed $\sim 10\,{\rm s}$ from SN1987A. This sets a limit on the coupling strength of the new particles. However, this constraint is not valid for arbitrarily strong couplings. For sufficiently large coupling strengths, the new particles would become trapped in the supernova core, preventing them from escaping and creating an exclusion window in the mass-coupling parameter space. Conversely, if the coupling is too weak, the particles would not be produced in sufficient quantities to impact the supernova dynamics.

This cooling argument has been widely applied to place limits on a variety of weakly interacting particles, including axions~\cite{Raffelt:1987yt, Dolan:2017osp, Chang:2018rso, Carenza:2019pxu, Carenza:2020cis,Dev:2020eam, Lucente:2020whw, Calore:2021klc, Caputo:2024oqc}, dark gauge bosons linked to new symmetries connected with the extension of the SM~\cite{Dent:2012mx, Kazanas:2014mca,Rrapaj:2015wgs,Chang:2016ntp,Mahoney:2017jqk, DeRocco:2019jti,Croon:2020lrf,Sung:2021swd,Caputo:2021eaa,Caputo:2021rux, Cerdeno:2023kqo,Akita:2023iwq,Cerdeno:2023kqo,Lai:2024mse}, and light scalars such as majorons that couple to neutrinos~\cite{Hardy:2024gwy,Brune:2018sab, Heurtier:2016otg,Chen:2022kal,Diamond:2023scc,Antel:2023hkf,Cappiello:2025tws}
. Furthermore, the decay of hypothetical bosonic particles into neutrinos~\cite{Fiorillo:2022cdq} or photons~\cite{Caputo:2022mah} within the core of a collapsing star could also alter the dynamics of the supernova, offering new avenues for probing the existence of these particles.
Recent refinements of this cooling argument have incorporated previously neglected physical effects, offering a more detailed framework for deriving constraints from updated supernova models~\cite{Caputo:2021rux,Caputo:2022rca}. However, it is important to note that in scenarios with significant self-interactions within the dark sector, these cooling constraints may no longer be applicable~\cite{Fiorillo:2024upk}.

Within the family of weakly interacting new particles which can be added to the SM, the dark photon, the gauge boson of a hidden U(1) symmetry kinetically mixed with the SM photon, has received considerable attention, both from a theoretical and experimental point of view~\cite{Holdom:1985ag, Bauer:2018onh,Fabbrichesi:2020wbt}. Such U(1) sectors frequently incorporate viable dark matter (DM) candidates, with the dark photon serving as the link between the visible and hidden states. 
As a result, dark photons are actively searched for in laboratory settings, including fixed-target experiments, meson decays, and colliders. Not only this, dark photons can have important astrophysical signatures. Depending on its mass $m_{A'}$ and mixing parameter $\epsilon$, dark photons can be copiously produced in the SN core and contribute to anomalous cooling. This has been used to exclude wide regions of the $(m_{A'},\epsilon)$ parameter space~\cite{An:2013yfc, Rrapaj:2015wgs, Chang:2016ntp, Caputo:2021eaa, Caputo:2025aac}. It is worth emphasizing that the results derived in this work are expected to remain qualitatively robust for a broader class of mediators beyond dark photons. The choice of dark photons in our analysis serves as a concrete and well-motivated benchmark scenario.

One important ingredient has been largely neglected in studies involving SN cooling: the fact that the progenitor star resides in a galactic DM halo. Over its lifetime, the star inevitably captures DM from its surroundings. These DM particles get captured gravitationally, lose energy by colliding with the nucleons, and finally settle inside the progenitor star. The capture may occur after a single or multiple collisions, depending on the strength of the DM interaction with the star's constituents and the DM mass. The capture rate for such a process has been calculated in detail for single \cite{spergel,gould1,gould2,gould3,silk:1985ax,krauss:1985aaa} and multiple scattering regimes \cite{1703.04043, Dasgupta_2019}. The presence of this DM population provides additional scattering targets for dark photons, thereby modifying their free-streaming length and, in turn, the cooling bounds. 

A preliminary discussion on this topic was conducted in \cite{Zhang:2014wra}, where the author discussed the impact of DM capture on the cooling rate of dark photons.
\MS{ The capture calculation employed there effectively assumes that the mediator mass is large compared to the typical momentum transfer involved in the scattering process, allowing the interaction to be treated as contact-like. This approximation is appropriate when $m_{A'}^2 \gg q^2$, where $q \sim \mu v_{\rm rel}$ denotes the momentum transfer and $\mu$ is the reduced mass of the scattering system. For GeV-scale dark matter scattering off nucleons, and typical capture velocities, $q\sim$ MeV. Consequently, for dark-photon masses below this scale, the momentum dependence of the propagator cannot be neglected. In this regime, the differential scattering cross section becomes strongly forward-peaked and the capture probability is modified significantly. The present work incorporates these light-mediator effects self-consistently and therefore extends the analysis into the low-$m_{A'}$ region where the heavy-mediator approximation ceases to be reliable.}

The interaction between DM and its associated mediator particle can significantly influence supernova cooling.
When the momentum transfer in a DM-nucleon collision is much smaller than the dark photon mass, the dynamics simplify and can be treated effectively as a contact interaction. In contrast, if the dark photon is light, the capture probability is altered in a non-trivial way, making the calculation more intricate. The goal of this work is to present a detailed and self-consistent analysis of how DM capture in SN progenitors alters the cooling constraints on dark photons. We systematically review the standard cooling calculation, discuss the capture and thermalisation of DM (including both annihilating and asymmetric cases), derive the modified free-streaming conditions, and compute the resulting luminosity. We then present numerical results showing the modifications to the exclusion regions in the dark photon parameter space. Our analysis demonstrates that previously excluded regions may reopen, and that the size of the effect depends sensitively on DM properties such as mass, annihilation rate, and self-interaction cross section.

Before proceeding, we emphasise that the SN cooling calculation employed in this work follows the standard analytic framework commonly used in the literature. More refined treatments of the SN1987A constraints incorporate effects such as in-medium photon dispersion, detailed radial temperature and density profiles, nuclear correlations, and many-body plasma interactions~\cite{Chang:2016ntp,Caputo:2021eaa,Hardy_2017}. While such improvements can quantitatively shift the cooling bounds, including all of them simultaneously can obscure the specific physical impact we aim to isolate here: the modification of the dark-photon free-streaming length due to the presence of captured DM. Since our goal is to demonstrate and characterise this effect in a controlled setting, we retain the minimal and transparent cooling prescription.Our objective is to isolate the effect of an additional opacity source provided by captured dark matter particles on SN cooling. Consequently, the exclusion regions presented in this work should be interpreted as illustrative of the alteration of the SN cooling bounds rather than definitive. A dedicated follow-up analysis, including realistic SN simulations and the impact of a thermal plasma, will be presented in future work. 

The paper is structured as follows. In Sec.\,\ref{sec:2}, we discuss the usual mechanism of anomalous SN cooling due to the existence of dark photons. Sec.\,\ref{sec:3} focuses on the mechanism of DM capture inside a supernova, expanding on the case of annihilating and non-annihilating DM. In Sec.\,\ref{sec:4}, we derive the new cooling constraints on dark photon in the case where DM has been captured in the SN core and discuss the revival of the dark photon parameter space. Sec.\,\ref{sec:5} concludes with the importance of our results and discusses possible future directions.

\section{SN cooling via emission of dark gauge boson}
\label{sec:2}
We consider a simple extension of the SM, extended by a dark photon $A'_\mu$ associated with a new U(1) gauge symmetry. The dark photon mixes with the SM photon with a kinetic mixing $\epsilon$. 
The interaction with the SM photon $A_\mu$ arises via kinetic mixing:
\begin{equation}
\mathcal{L}_{A-A'} \supset -\frac{1}{4} F_{\mu\nu}F^{\mu\nu} - \frac{1}{4} F'_{\mu\nu}F'^{\mu\nu} - \frac{\epsilon}{2} F_{\mu\nu} F'^{\mu\nu} + \frac{1}{2} m_{A'}^2 A'_\mu A'^\mu \,,
\label{eq:Lag}
\end{equation}
where  $m_{A'}$ denotes the masses of the dark photon. We stay agnostic about how the mass is generated in this specific study. The production of these dark photons, through kinetic mixing with the SM photon, and their subsequent emission, can accelerate the cooling of the proto-neutron star and thus shorten the observed duration of the neutrino signal. Observational consistency with the $\sim 10$ second neutrino burst from SN1987A imposes stringent constraints on the kinetic mixing parameter $\epsilon$ across a broad range of dark photon masses.

Inside the SN core,  with temperatures $T \sim 30$ MeV and densities $\rho \sim 3\times 10^{14}$ g/cm$^3$, dark photons can be produced primarily via proton-proton bremsstrahlung, $p + p \rightarrow p + p + A'$, with subdominant contributions coming from electron-positron annihilation and plasmon decays. The cross-section for the process can be estimated as~\cite{Zhang:2014wra}
\begin{equation}
    \sigma_{pA'} (T)\simeq \frac{6\epsilon^2 \alpha m_p T}{\pi^2 m_\pi^4}\,,
    \label{eq:sigmaPA}
\end{equation}
where $\alpha$ is the electromagnetic fine structure constant, $m_p$ and $m_\pi$ are the proton and pion masses respectively, and $T$ is the temperature of the medium. 

The fate of the produced dark photons depends on their mean free path, $\lambda_p \equiv 1/(n_p\, \sigma_{pA'})$. Two regimes are important. 
\begin{itemize}
\item \textbf{Free-streaming regime (Volume emission):} For small $\epsilon$, dark photons escape unimpeded if $\lambda_p \gg R_{\rm core}$, where $R_{\rm core} \sim 10$ km is the PNS core radius. 
In this case, the total luminosity is given by volume emission:
\begin{equation}
L_{A'}^{\rm vol} \simeq V_c \, n_p^2 \, T_c \, \sigma_{pA'} \, \exp\left(-\frac{m_{A'}}{T_c}\right)\,\exp\left(-\Gamma_{\rm decay}\,R_{\rm core}\right)\,,
    \label{eq:LAvol}
\end{equation}
where $V_c=4\pi R_{\rm core}^3/3$ is the volume of the PNS core, $n_p\equiv N_p/V_c = 1.2 \times 10^{38}\,{\rm cm}^{-3}$ is the number density of protons in the core, and $T_c=30\,{\rm MeV}$ is the temperature of the SN core. For $m_{A^\prime}$ greater than the core temperatures of the neutron star, a Boltzmann suppression factor of $e^{-m_{A^\prime}/T_c}$ is taken into account for calculating the final luminosity.
Hence, the number density falls off with decreasing $m_{A^\prime}$, which in turn leads to a decrease in the final luminosity. The last factor denotes an additional reduction of the luminosity due to the two-body decay of $A^\prime$ to SM leptons \cite{dent2012,Kazanas_2015}. 
Here $\Gamma_{\rm decay}$ denotes the decay width of the dark photon into SM states. For $m_{A'} \gtrsim 2 m_e$, this is approximately given by
\begin{equation}
    \Gamma\,\left(A^\prime \rightarrow e^+\,e^- \right)=\dfrac{1}{3}\,\alpha\epsilon^2 m_{A^\prime}\left(1-\dfrac{4m_e^2}{m_{A^\prime}^2}\right)^{1/2}\left(1+\dfrac{2m_e^2}{m_{A^\prime}^2}\right)\dfrac{m_{A^\prime}}{E_{A^\prime}}
\end{equation}

If the $L_{A'}^{\rm vol}$ exceeds the typical expected neutrino luminosity $L_0 = 3\times 10^{52} \rm{erg}/\rm{s}$~\cite{Raffelt:1987yt}, then the cooling timescale becomes too short, conflicting with the observed neutrino signal. This sets an upper limit on the allowed $\epsilon$ in this regime.

\item \textbf{Trapped regime (Surface emission):}  The volume emission cannot go on indefinitely. As $\epsilon$ increases, the interaction rate grows, eventually leading $\lambda_p \ll R_{\rm core}$.  In this limit, dark photons thermalise and get trapped inside a ``dark photosphere'', much like the neutrinosphere.
Energy loss is then controlled by blackbody-like emission from the surface of the effective dark photosphere located at the decoupling radius $R_{\rm{dec}}$. The radius of decoupling is determined from the condition that the optical depth $\tau$~\cite{Janka:2006fh}. 
\begin{equation}
    \tau \equiv \int_{R_{\rm dec}}^{\infty}\, dr \, n_p(r)\,  \sigma_{pA'} (T(r)) = \frac{2}{3}\,,
\end{equation}
where the radial dependence enters through the temperature profile.
The luminosity of $A'$ through blackbody radiation from the photosphere is governed by the Stefan-Boltzmann law,
\begin{equation}
    L_{A'}^{\rm{surf}} =  4\pi R_{\rm{dec}}^2\,\sigma_B\,T_{\rm{dec}}^4\,\exp\left(-\frac{m_{A'}}{T_{\rm dec}}\right)\,,
    \label{eq:LAsurf}
\end{equation} 
where $T_{\mathrm{dec}}$ is the local temperature at the decoupling surface and $\sigma_B= 3\pi^2/120$ is the Stefan-Boltzmann constant. If $ L_{A'}^{\rm{surf}} > L_0$, then the SN cooling becomes efficient again. This leads to a ceiling in the exclusion limit for $\epsilon$. 

\end{itemize}

To study the cooling constraint, we consider the following parametric profile for the temperature and number density, respectively~\cite{Zhang:2014wra},
\begin{equation}
     T(r)=\begin{cases} 
      T_c & r\leq R_{\rm core}\, \\
  T_c \left(\frac{r}{R_{\rm core}}\right)^{-5/3} & r > R_{\rm core}\, \\
   \end{cases}
   \label{eq:TempProf}
\end{equation} 
and 
\begin{equation}
    n_p(r)=\begin{cases} 
      n_p & r\leq R_{\rm core}\, \\
  n_p \left(\frac{r}{R_{\rm core}}\right)^{-5} & r > R_{\rm core}\,\,. \\
   \end{cases}
   \label{eq:numdenProf}
\end{equation} 
Note that these parametric profiles are considered for ease of computation to demonstrate the effect of DM capture in this scenario. A more realistic analysis, with temperature and density profiles taken from a simulation, will be performed in the future. 

\begin{figure}[!t]
\includegraphics[width=0.45\linewidth, height=0.35\linewidth]{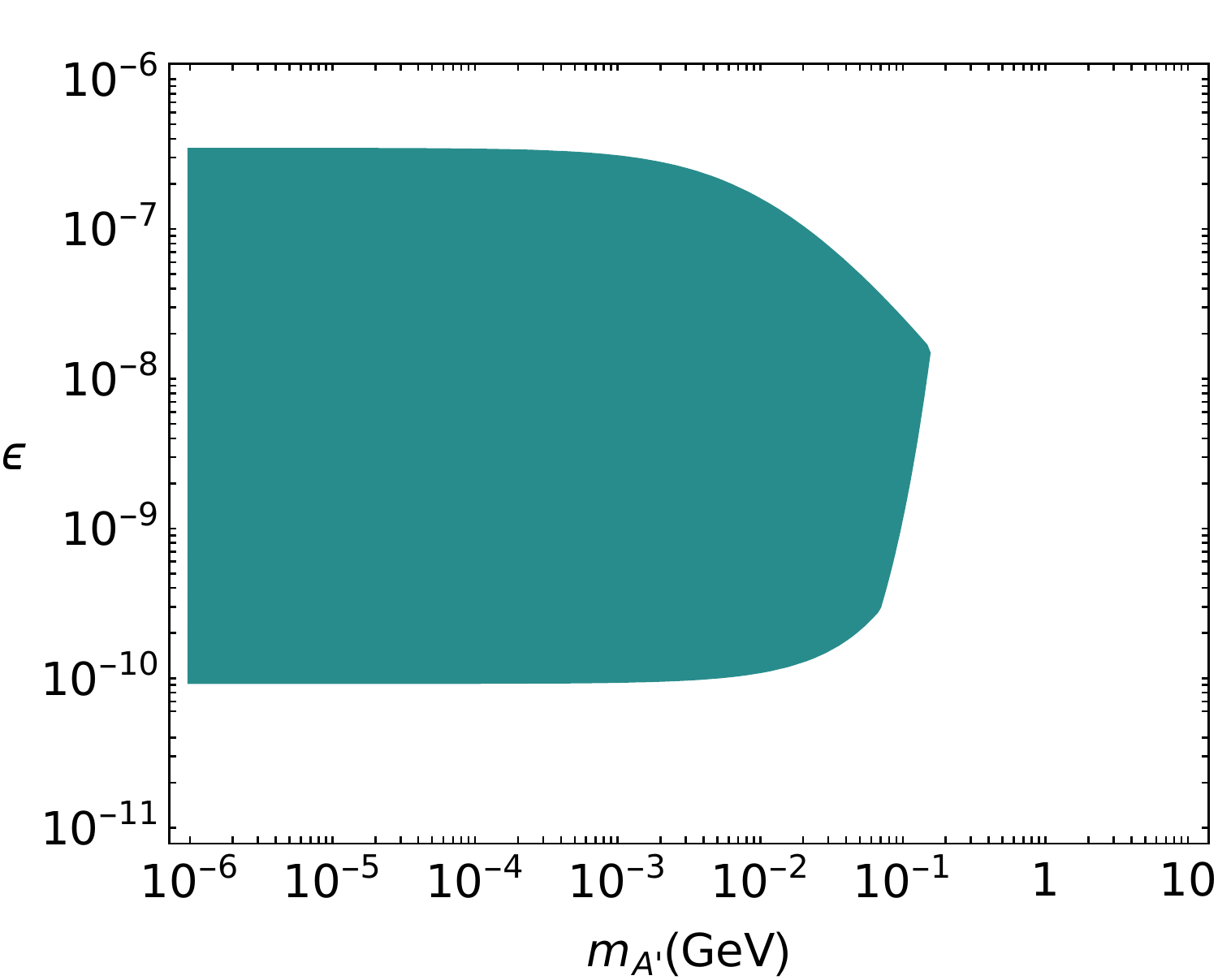}
    \caption{The standard SN1987A cooling bound forms an exclusion band in the $(m_{A'},\epsilon)$ plane. For very small kinetic mixing, dark photons are produced too weakly to alter the cooling. For very large mixing, frequent scatterings trap and thermalise the $A'$, suppressing energy loss. Only in the intermediate regime do $A'$ particles free-stream out and overcool the proto-neutron star, leading to the excluded region.}
     \label{fig:SNCooling}
 \end{figure}

Fig.\,(\ref{fig:SNCooling}) shows the interplay of these two regimes, producing the characteristic SN cooling exclusion contour. For very small kinetic mixing, $A'$ production is too suppressed to modify the cooling rate, while for very large mixing, the $A'$ particles scatter efficiently and become trapped, thermalising with the stellar medium. 
Only in the intermediate regime do $A'$ particles free-stream out and remove energy efficiently, leading to excessive cooling. 
This results in the characteristic excluded band in Fig.~(\ref{fig:SNCooling}), bounded on one side by underproduction and on the other by trapping.

The calculation presented above is intuitive and does not include non-trivial modifications of the photon propagator in the dense, hot medium of the SN~\cite{Chang:2016ntp}. Such finite-density plasma effects have been shown to suppress the in-medium kinetic mixing for the dark photon frequency below the plasma frequency. This suppression reduces both production and absorption rates for low-energy dark photons, thereby loosening the bounds. Additional improvements include realistic radial profiles of temperature and density in the SN core~\cite{Caputo:2021rux}. Broadly, these improvements yield more accurate constraints, typically weakening the bounds relative to earlier estimates. In this work we do not attempt to incorporate these refinements. Our goal, instead, is to isolate the impact of an additional opacity source arising from captured dark matter. 
Importantly, the mechanism we identify depends primarily on the relative contribution of different scattering channels to the total opacity. Plasma effects effectively rescale the baryonic contribution to the opacity but do not eliminate the possibility that the DM contribution dominates. Therefore, while the precise location and shape of the exclusion contours are expected to change in a more complete treatment, the qualitative effect of dark matter-induced trapping remains robust. A detailed study incorporating in-medium effects and realistic supernova profiles is left for future work.

\section{Captured dark matter inside Supernovae}
\label{sec:3}
Stars embedded in a galactic halo gravitationally capture DM via scatterings with the constituent nucleons. An incoming DM particle with velocity $u$ relative to the star scatters off a nucleon, loses kinetic energy, and becomes bound if its post-collision velocity is below the escape velocity $v_{\rm esc}$. After repeated scatterings, captured DM thermalises and settles inside the star. Hence, over the lifetime of the star, \MS{prior to core collapse}, we expect a considerable buildup of DM concentration inside the celestial object. \MS{The capture rates discussed in this section depend on progenitor properties such as the stellar radius, escape velocity, density profile, and lifetime. In contrast, the cooling calculation discussed in Sec.~\ref{sec:2} applies to the proto-neutron star formed after the collapse of the stellar core and is characterised by the core temperature and density.}

Note that, in this work,  we do not model the full time-dependent evolution of the stellar profile after the DM capture. Instead, we assume that the DM population accumulated during the progenitor phase is retained during the formation of the proto-neutron core after the collapse. This approximation allows us to estimate the impact of the captured dark matter on the post-collapse cooling phase, while keeping the analysis analytically tractable. The conditions under which this can be feasible is being investigated in a follow-up work. 

The captured DM alters the landscape of the cooling mechanism dramatically. \MS{The captured DM population is characterised by a local number density $n_\chi(r)$.}
This happens primarily because the presence of DM provides an additional scattering channel for $A'$, with a mean free path given by $\lambda_\chi \equiv 1/(n_\chi\, \sigma_{\chi\,A'})$, where \MS{$\sigma_{\chi A'}$ indicates the DM-dark photon scattering cross-section}. Hence, the total inverse mean free path is modified as
\begin{equation}
\lambda_{\rm eff}^{-1}(r) = n_p(r)\,\sigma_{pA'} + n_\chi(r)\,\sigma_{\chi A'},
\label{eq:Lambdaeff}
\end{equation}

Whether DM affects the cooling depends on whether the DM-dark photon scattering contributes significantly to the opacity. This is determined by comparing $n_\chi(r)\sigma_{\chi A'}$ with the baryonic contribution
$n_p(r)\sigma_{pA'}$.
Two cases may arise. If $\lambda_\chi > \lambda_p$, interaction of $A'$ with nucleons dominates, the cooling mechanism remains the same as described in section~\ref{sec:2}. On the other hand, if $\lambda_\chi < \lambda_p$, the interaction of $A'$ with DM dominates. This can lead to the formation of the photosphere, a surface of last scattering of $A'$ with DM, with a radius $r_{A'}$, defined by 
\begin{equation}
\int_{r_{A'}}^\infty dr \, \left[ n_p(r)\sigma_{p A'} + n_\chi(r)\sigma_{\chi A'}\right] = \frac{2}{3}.
\end{equation}
If $r_{A'}<R_{\rm core}$, the luminosity is due to the emission from the annular volume between the core surface and the surface of the $r_{A^\prime}$ sphere and is $\propto \left(R_{\rm core}^3-r_{A^\prime}^3\right)$. With an increasing amount of captured dark matter particles, $r_{A^\prime}$ increases as well. Finally, when $r_{A'}\gtrsim R_{\rm core}$, all emission originates from the surface at $r_{A'}$, and there is no volume emission. This interplay between volume and surface emission can change the allowed parameter space drastically. We will expand on these discussions in the following sections.

To demonstrate this point, we consider a fermionic DM particle $\chi$ interacting with the dark photon. The Lagrangian, including DM interactions, is given by
\begin{equation}
    \mathcal{L} \supset \mathcal{L}_{\rm SM} - \frac{1}{4} F'_{\mu\nu}F'^{\mu\nu}  - \frac{\epsilon}{2} F_{\mu \nu} F'^{\mu \nu} + \bar{\chi} i\gamma^\mu \left(\partial_\mu - i e^\prime A'_\mu \right)\chi + m_\chi \bar{\chi} \chi + m_{A'}^2 A'_\mu A'^{\mu}\,,
    \label{eq:Lag2}
\end{equation}
where $m_\chi$ is the mass of DM and $e'$ is the coupling of DM with $A'$. As before, we stay agnostic about how the DM mass is generated. We discuss the capture process for annihilating DM and non-annihilating (asymmetric) DM, respectively.
\subsection{Annihilating Dark matter}
\label{sec:anndm}
The time evolution of the number of annihilating DM particles captured inside the star is given by
\begin{equation}
    \dfrac{dN_\chi}{dt}=C_0+C_{\rm s}\,N_\chi - C_{\rm e} \, N_\chi - C_{\rm se} N_\chi^2 - C_{\rm ann}\,N_\chi^2\, .
\end{equation}
Here, $C_0$ denotes the capture rate of DM through scattering with nucleons, 
$C_{\rm s}$ represents the self-capture rate arising from collisions with previously captured DM particles, 
$C_{\rm e}$ ($C_{\rm se}$) corresponds to the rate of evaporation (self-evaporation), 
and $C_{\rm ann}$ is the annihilation rate of the captured DM population into other particles. 

For very light DM, the typical kinetic energy can become comparable to the star’s internal thermal energy. 
In such cases, thermal collisions may accelerate DM particles to velocities exceeding the local escape velocity, 
allowing them to leave the gravitational potential well---this process gives rise to the $C_{\rm e} N_\chi$ term. 
Similarly, when two captured DM particles scatter off each other, one of them may acquire sufficient energy to escape. 
The required energy for such collisions again originates from the stellar thermal bath. 
Since this process involves two DM particles, it contributes to the self-evaporation term $C_{\rm se} N_\chi^2$. 
These evaporation effects are particularly relevant for light, sub-GeV DM candidates. 

In the present work, we focus on heavier DM particles, for which both evaporation and self-evaporation are negligible. 
Consequently, the evolution equation simplifies to
\begin{equation}
    \frac{dN_\chi}{dt} = C_0 + C_{\rm s} N_\chi - C_{\rm ann} N_\chi^2\, .
    \label{eq:annDM1}
\end{equation}

\MS{The DM nucleon capture rate is given by~\cite{Zentner:2009is} 
\begin{eqnarray}
    C_0 & = & \sqrt{\frac{3}{2}}\,N_p\left(\frac{\rho_\chi}{m_\chi}\right)\langle \phi \rangle \,\frac{{\rm Erf}(\eta)}{\eta}\frac{v_{\rm esc}^2}{v_\chi}\,{\rm Min}\left(\sigma_{\rm sat},\sigma_{\chi p} \right)\,,
\end{eqnarray}
where $N_p$ the total number of nucleons, $v_{\rm esc}$ is the escape velocity at the surface, $v_\chi$ is the DM local velocity dispersion in the halo, and $\sigma_{\chi p} $ is the DM-nucleon scattering cross section. Here $\langle \phi \rangle \sim 5 $ is a dimensionless parameter characterising the average potential due to a Sun-like star, and $\eta = 1.5\,(v_\odot/v_\chi)^2$ is another dimensionless quantity characterising the velocity of a Sun-like star through the halo.}

At early times, the DM population inside the star grows primarily through capture via collisions with nucleons. 
The capture rate $C_0$ scales with $\sigma_{\chi p}$, which, however, cannot increase indefinitely, and is limited by the radius of the star, $R_\star$. Once every incoming DM particle interacts at least once while passing through the star, the rate saturates at the geometric limit
\begin{equation}
    \sigma_{\rm sat} = \frac{\pi R_\star^2}{N_p}\,,
\end{equation}
where $R_\star$ is the stellar radius. For our calculation, we use the radius of the progenitor star, $R_\star\sim 8 R_{\odot}$. 

\MS{The self-scattering capture rate is given by
\begin{equation}
    C_s =\sqrt{\frac{3}{2}}\,n_\chi \langle \phi \rangle \,\frac{{\rm Erf}(\eta)}{\eta}\frac{v_{\rm esc}^2}{v_\chi}\,{\rm Min}\left(\sigma^{\rm sat}_s,\sigma_s \right),
\end{equation}
where $\sigma_s$ is the DM self-interaction cross-section, and $\sigma^{\rm sat}_s$ is the maximum allowed self-interaction cross-section, defined in Eq.\,\ref{eq:selfcsmax} in the next subsection.
}

Captured DM particles gradually lose kinetic energy through successive scatterings and eventually thermalise with the stellar medium. 
%
%
\MS{ The thermal radius relevant for the post-collapse cooling calculation is obtained using the thermodynamic conditions of the proto-neutron-star core. Once the collapse has occurred, the previously captured DM population is assumed to thermalise within a sphere of radius,
\begin{equation}
    r_{\rm th} = \left( \frac{9\,T}{4\pi\,G\,m_\chi\,\rho} \right)^{1/2}\,.
    \label{eq:rtherm}
\end{equation}
For a core density of $3\times10^{14}\,\rm gm \,\rm cm^{-3}$ and a core temperature of 30 MeV, we find $r_{\rm th}\simeq 10\,{\rm km}/\sqrt{m_\chi/(1\,{\rm GeV})}\,$.}

\MS{The annihilation rate $C_{\rm ann}$ is given by
\begin{equation}
   C_{\rm ann} = \frac{\langle\sigma_{\rm ann}\, v_\chi \rangle}{\dfrac{4}{3}\pi\,r_{\rm th}^3}\,
\end{equation}
where $\langle\sigma_{\rm ann}\, v_\chi \rangle$ is the thermally averaged annihilation cross-section. For a detailed discussion of the derivation of these capture rates, see~\cite{Zentner:2009is,Chen:2014hha}.}

\subsubsection*{(i) Regime before self-interaction saturation ($t < t_{\rm crit}$)}
Initially, the DM number increases linearly due to the capture term $C_0$. As $N_\chi$ builds up, self-capture becomes important through the term $C_{\rm s} N_\chi$. The time $t_{\rm crit}$ when this transition occurs is defined by the condition $C_0 \simeq C_{\rm s}\,N_\chi(t_{\rm crit})$, yielding
\begin{equation}
    t_{\rm crit}=\dfrac{1}{C_{\rm s}}\ln\Big(1+ \dfrac{\pi r_{\rm th}^2}{\sigma_s}\dfrac{C_{\rm s}}{C_0}\Big)\,,
\end{equation}
where $\sigma_{\rm s}$ is the DM self-interaction cross section.

For $t < t_{\rm crit}$, Eq.\,(\ref{eq:annDM1}) governs the evolution. The general solution is
\begin{equation}
N_\chi(t) = C_0\,\tau_0\,\frac{\tanh\left(\kappa\, t / \tau_0\right)}{\kappa - \frac{1}{2} C_{\rm s}\, \tau_0 \,\tanh\left(\kappa\, t / \tau_0\right)}\,,
\label{eq:gensoln1}
\end{equation}
where
$\tau_0 = 1/\sqrt{C_{\rm ann}\,C_0}$ and $\kappa = \sqrt{1 + \left( (C_{\rm s}\,\tau_0)/2 \right)^2}$.

Equilibrium is reached when the RHS of Eq.\,(\ref{eq:annDM1})  vanishes. This yields $\tanh(\kappa t / \tau_0) \simeq 1$, corresponding to a timescale $t_{\rm eq,1} \simeq \tau_0/\kappa$. The equilibrium number of captured DM particles in this regime is
\begin{equation}
   N_\chi^{\rm eq,\,1}= \dfrac{C_0\,\tau_0}{\kappa - \dfrac{C_{\rm s}\, \tau_0}{2} }\,.
   \label{eq:dmann_sat_1}
\end{equation}
If the system reaches equilibrium before $t_{\rm crit}$, $N_\chi$ saturates to this value.

\subsubsection*{(ii) Regime after self-interaction saturation ($t > t_{\rm crit}$)}

When the self-interaction cross section approaches the geometric limit, self-capture saturates. Since most of the captured DM resides within the thermal radius $r_{\rm th}$, the maximum allowed self-interaction cross section is given by 
\begin{equation}
    \sigma_{\rm s}^{\rm sat} = \frac{\pi r_{\rm th}^2}{N_\chi(t_{\rm crit})}.
    \label{eq:selfcsmax}
\end{equation}
\MS{We emphasise that $t_{\rm crit}$ does not correspond to saturation of the total DM population. Rather, it marks the onset of geometric saturation of the self-capture process. At this stage the effective self-capture cross section reaches the geometric limit, and further increases in the number of captured particles no longer enhance the self-capture probability. The DM population itself may continue to evolve substantially after $t_{\rm crit}$, including periods of exponential growth, before eventually approaching its late-time equilibrium value. Consequently, $N_\chi(t_{\rm crit})$ should not be interpreted as the final saturated population of DM inside the star.}

\MS{For $t>t_{\rm crit}$}, the differential equation Eq.\,(\ref{eq:annDM1}) will have no linear dependence on $N_\chi$. The rate equation takes the modified form,
\begin{equation}
    \dfrac{dN_\chi}{dt}=C_0+C_{\rm s}^{\rm sat} - C_{\rm ann}\,N_\chi^{2} \,,\hspace{0.5cm} \left(t > t_{\rm crit}\right)\,,
    \label{eq:annDM2}
\end{equation}
where, $C_{\rm s}^{\rm sat}$ is the self capture rate with $\sigma_s$ replaced by $\pi\,r_{\rm th}^2/N(t_{\rm crit})$.

The general solution to Eq.\,(\ref{eq:annDM2}) is
\begin{equation}
    N_\chi(t)= \sqrt{\dfrac{C_0+C_{\rm s}^{\rm sat}}{C_{\rm ann}}}\tanh\left(t\,\sqrt{C_{\rm ann}\,\left(C_0+C_{\rm s}^{\rm sat}\right)}\right)\,.
    \label{eq:gensol2}
\end{equation}
In this case, equilibrium is attained when $t_{\rm eq,\,2} \sim \left(C_{\rm ann}\,\left(C_0+C_{\rm s}^{\rm sat}\right)\right)^{-1/2} $ and the corresponding equilibrium number is given by
\begin{equation}
    N_\chi^{\rm eq,\,2}= \sqrt{\dfrac{C_0+C_{\rm s}^{\rm sat}}{C_{\rm ann}}}\,.
    \label{eq:eq2}
\end{equation}
These quantities are relevant only if the system reaches $t_{\rm crit}$ within the stellar lifetime.

In this work, \MS{we use the progenitor lifetime, $t_\star \sim 3.75~{\rm Myr}$, as an order-of-magnitude estimate of the available capture time.}
Typically, this timescale is much longer than all other relevant timescales in the problem. In particular, we consider the regime in which the system has sufficient time to either reach equilibrium or enter the self-interaction-dominated phase. This allows us to focus on the late-time behaviour of the DM population.

\MS{The evolution of the captured DM number is controlled by two key timescales: the equilibration time ($t_{\rm eq}$) when $dN_\chi/dt$ vanishes and the critical time ($t_{\rm crit}$) at which the self-capture rate starts to depend on the corresponding self-interaction saturation cross section ($\sigma^{\rm sat}_s$). The relative ordering of these timescales determines the qualitative behaviour of the system. Broadly, two possibilities arise:
\begin{itemize}
    \item[(i)] if $t_{\rm eq,1} < t_{\rm crit}$, the evolution is entirely governed by Eq.\,(\ref{eq:annDM1}), else
    \item[(ii)] if $t_{\rm crit} < t_{\rm eq,1}$, the system eventually transitions to the self-interaction dominated regime described by Eq.\,(\ref{eq:annDM2}).
\end{itemize}
These possibilities lead to the following evolutionary stages:}

\begin{enumerate}
 \item \textbf{Early linear growth}: \\
   \MS{ At early times, the DM population grows due to capture by nucleons. The evolution is governed mostly by the first term of Eq.\,(\ref{eq:annDM1}) and the growth is linear in time.}

  \item \textbf{Equilibration before $t_{\rm crit}$} ($\sigma_{\chi\chi} < \sigma_{\rm s}^{\rm sat}$): \\
  \MS{  As dark matter particles begin to accumulate, self-capture starts to become important. The linear growth is followed by an exponential one controlled by the term $C_s\,N_\chi$. Finally, the DM population saturates at the value given by Eq.\,(\ref{eq:dmann_sat_1}). Beyond this point, the number of captured DM particles remain constant.}

  \item \textbf{Equilibration after $t_{\rm crit}$} ($\sigma_{\chi\chi} > \sigma_{\rm s}^{\rm sat}$): \\
   \MS{ On the other hand, if the system reaches criticality before $t_{\rm eq,\, 1}$, the exponential growth due to self capture is replaced by a linear one as dictated by the second term of Eq.\,(\ref{eq:annDM2}). In this phase, the DM population continues to grow until it approaches a new equilibrium determined by the balance between capture, saturated self-capture, and annihilation. At $t_{\rm eq,\,2}$, the system reaches its final equilibrium configuration, with $N_\chi$ saturating to the value given in Eq.\,(\ref{eq:eq2}).}
\end{enumerate}

\begin{figure}[!t]
    \centering
    \includegraphics[width=0.45\linewidth]{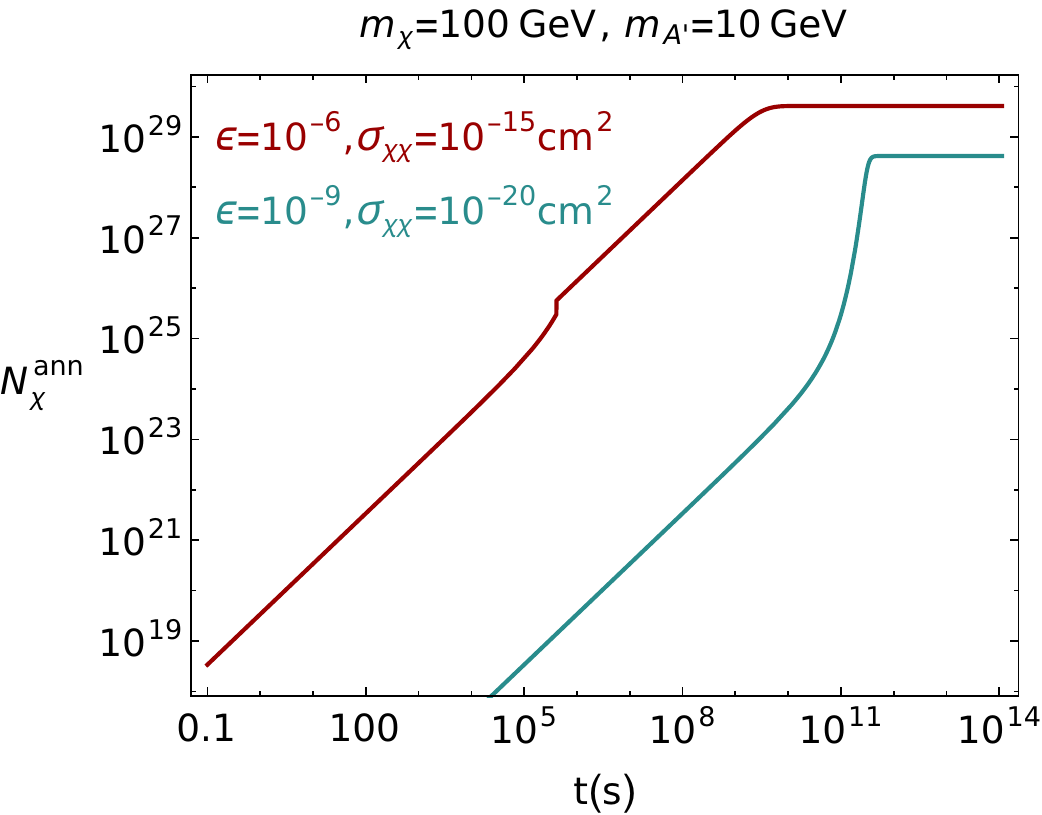}
    \caption{Time evolution of the total number of annihilating DM captured inside the neutron star. The teal line denotes a benchmark where evolution is governed by Eq.\,(\ref{eq:annDM1}) and DM number grows by passing through a linear phase, followed by an exponential phase, before saturating to a plateau given by Eq.\,(\ref{eq:dmann_sat_1}). The red line denotes a benchmark where the system passes through subsequent phases of linear growth, with an interim period of exponential growth, before reaching saturation given by Eq.\,(\ref{eq:eq2}). Here we used $v_{\rm esc} \sim 1000\,\rm km\,s^{-1}$ and local dark matter density to be 0.4 GeV$\,\rm cm^{-3}$.}
    \label{fig:Nchi_t_ann}
\end{figure}

The resulting time evolution of $N_\chi$ for two representative parameter sets is shown in Fig.\,(\ref{fig:Nchi_t_ann}). 
For $\sigma_{\chi\chi} < \sigma_{\rm s}^{\rm sat}$, $N_\chi$ exhibits an initial linear growth, followed by an exponential increase before reaching equilibrium, as described by Eq.\,(\ref{eq:gensoln1}). This behaviour is illustrated by the teal curve corresponding to $\epsilon=10^{-9}$ and $\sigma_{\chi\chi}=10^{-20}\,{\rm cm}^2$.
In contrast, when $\sigma_{\chi\chi} > \sigma^{\rm sat}_{s}$, the evolution follows Eq.\,(\ref{eq:annDM2}), showing two linear growth phases separated by an intermediate exponential phase before saturation. This is shown by the red curve corresponding to $\epsilon=10^{-6}$ and $\sigma_{\chi\chi}=10^{-15}\,{\rm cm}^2$.

\subsection{Non-annihilating dark matter}
\label{sec:nonanndm}
\begin{figure}[!t]
    \centering
    \includegraphics[width=0.45\linewidth]{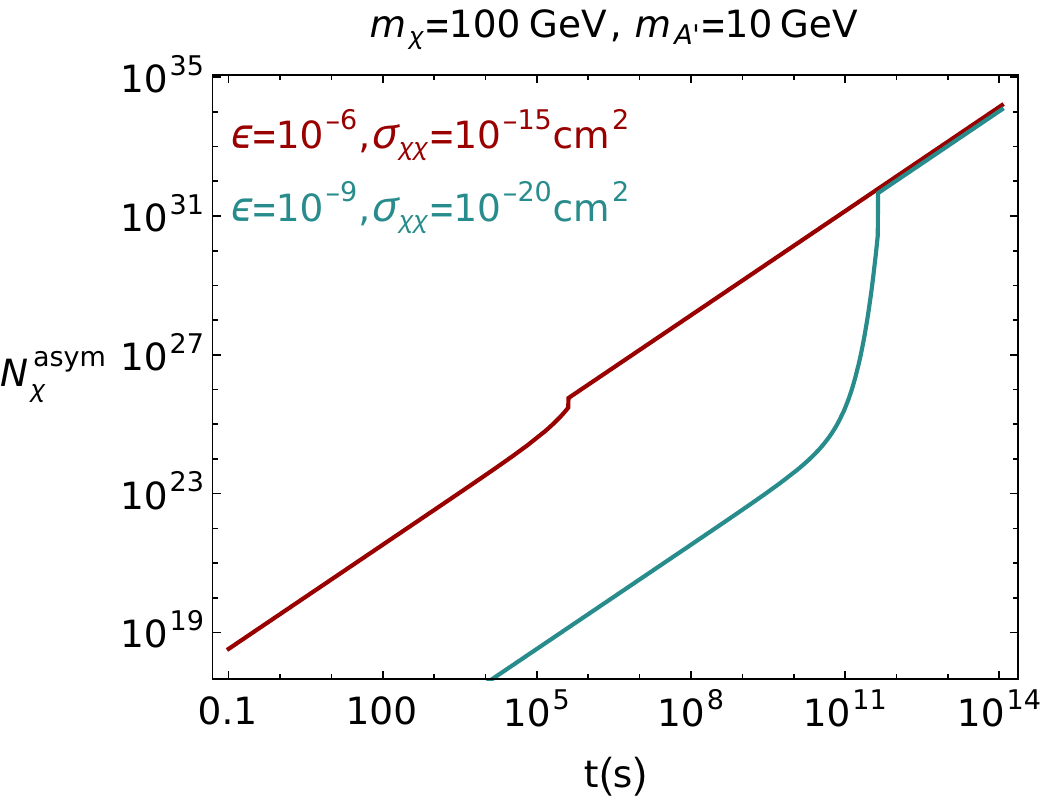}
    \caption{Time evolution of the total number of non-annihilating DM captured inside the neutron star. The teal line denotes a benchmark where the system passes through subsequent phases of linear and exponential growth, as described by Eq.\,(\ref{eq:assdm_1}). The red line, on the other hand, denotes a benchmark where evolution is governed by Eq.\,(\ref{eq:assdm_2}) and DM number grows by passing through two subsequent linear phases. The number of DM particles at large times coincides for these two cases.}
    \label{fig:Nchi_t_ass}
\end{figure}

Non-annihilating or asymmetric DM cannot deplete its population through mutual scattering and collisions among itself. Therefore, the $C_{\rm ann}$ term in Eq.\,(\ref{eq:annDM1}) or Eq.\,(\ref{eq:annDM2}) is absent. Consequently, the RHS of the evolution equation never vanishes, implying that the number of captured DM particles continues to increase over time. In this case, there exists only a single relevant timescale, namely $t_{\rm crit}$, since an equilibrium time no longer exists. 
The time evolution of the DM population is therefore governed by
\begin{eqnarray}
    \dfrac{dN_\chi}{dt}=C_0+C_{\rm s}\,N_\chi\,, \quad \left(t < t_{\rm crit}\right)\,,\label{eq:assdm_1} \,\, \text{and} \\
    \dfrac{dN_\chi}{dt}=C_0+C_{\rm s}^{\rm sat}\,, \quad \left(t > t_{\rm crit}\right)\,.
    \label{eq:assdm_2}
\end{eqnarray}
Initially, the number of captured DM  particles grows linearly with time, driven by the capture rate $C_0$. As the DM population builds up and $ C_{\rm s} \, N_\chi$ becomes significant, the growth becomes exponential. After the system reaches $t_{\rm crit}$, the self-capture rate saturates, and the growth becomes linear once again, but with a different slope than in the initial phase. The time evolution of $N_\chi$ for two representative parameter sets is illustrated in Fig.\,(\ref{fig:Nchi_t_ass}). The evolution is similar to Fig.\,(\ref{fig:Nchi_t_ann}) with the major difference being the absence of an equilibrium number density. This is an important property of asymmetric DM, where accretion can continue till the lifetime of the star.

Once the total amount of captured DM is determined, we can compute its spatial distribution within the star. Assuming that the DM velocities follow a Maxwell-Boltzmann distribution and that the stellar density remains approximately constant, the radial number density profile takes the analytic form
\begin{equation}
    n_\chi(r) =  \dfrac{N_\chi}{\pi^{3/2}\,r_{\rm th}^3}\,\exp\left(-\dfrac{r^2}{r_{\rm th}^2}\right)\,,
    \label{eq:DMdist}
\end{equation}
where $r_{\rm th}$ is the thermal radius of the DM cloud, as shown in Eq.\,(\ref{eq:rtherm}). 
This distribution will be used in the following section to evaluate the free-streaming length of the dark photon $A'$ as it propagates through and interacts with the accumulated DM population.

\subsection{Capture rates and light mediators}
In the absence of DM, regions of the parameter space corresponding to sub-GeV dark photons ($A'$) can be strongly constrained by SN cooling arguments. In this work, we focus on heavy DM particles of mass $\mathcal{O}(100\,{\rm GeV})$. 
The capture of such DM within the stellar core proceeds primarily through its elastic collisions with nucleons, mediated by a light dark photon $A'$ with a mass lower than the DM mass. Accordingly, the capture rate calculation incorporates the effect of these light mediators. We follow the formalism of \cite{Dasgupta_2020} to determine the capture rate of DM via its scattering with nucleons as well as with previously captured DM particles. The different capture rates are given by 
\begin{eqnarray}
    C_0 &=& \dfrac{\rho_\chi}{m_\chi}\int \dfrac{f(u)}{u}\,\left(u^2+v_{\rm esc}^2\right)\,du\,g_1(u)\, \rm Min\left[\sigma_{\chi p},\sigma^{\rm sat}\right],\\
    C_{\rm s} &=& \dfrac{\rho_\chi}{m_\chi}\int \dfrac{f(u)}{u}\,\left(u^2+v_{\rm esc}^2\right)\,du\,g_1^{\rm self}(u)\, \rm Min\left[\sigma_{\chi\chi},\sigma^{\rm sat}_{\chi\chi}\right],\\
    C_{\rm ann} &=& \dfrac{\langle\sigma_{\rm ann}\, v\rangle}{\dfrac{4}{3}\pi\,r_{\rm th}^3}\,,
    \label{eqn:Caprates}
\end{eqnarray}
where $\sigma_{\chi p}$, $\sigma_{\chi\chi}$, and $\sigma_{\rm ann}$ denote the DM-nucleon, DM self-interaction, and annihilation cross sections, respectively. In the non-relativistic approximation, the differential cross section of DM-nucleon and DM self-interactions can be well described by the Born scattering formula \cite{Petraki_2015,Petraki_2017}, leading to an effective Yukawa-type potential. Within our simplified model, these cross sections can be expressed in terms of the underlying parameters as
\begin{eqnarray}
\sigma_{\chi p} &=& \int_{\Delta\Omega}\dfrac{d\sigma_{\chi p}}{d\Omega}\,d\Omega\,, \quad \text{where,}\nonumber \\
\dfrac{d\sigma_{\chi p}}{d\Omega} &=& \dfrac{4\,\mu^2\, \epsilon^2}{(4\,\mu^2\,v_{\rm rel}^2\,\sin^2(\theta/2)+m_{A^\prime}^2)^2}\,.
\end{eqnarray}
Here, $\mu$ is the reduced mass of the DM and target particle system, $\theta$ is the scattering angle in the centre of mass frame and $v_{\rm rel}$ is the relative velocity of the incoming DM and the target particle. The annihilation cross section $\sigma_{\rm ann}$ is given by
\begin{eqnarray}
    \sigma_{\rm ann} = \dfrac{\pi\,\alpha^{\prime\,2}}{m_\chi^2}\left(\sqrt{1-\dfrac{m_{A^\prime}^2}{m_\chi^2}}\right)\,\Theta\left(m_\chi > m_{A^\prime}\right)\,,
\end{eqnarray}
where $\alpha^{\prime\,2}=e^{\prime\,2}/4\pi$ and $\Theta$ is the Heaviside theta function.

Returning to the capture rate expressions, $f(u)$ represents the Maxwell--Boltzmann velocity distribution of DM particles in the galactic halo, while $\rho_\chi$ and $m_\chi$ are the local DM density and DM mass, respectively. The functions $g_1(u)$ and $g_1^{\rm self}(u)$ quantify the probability that the velocity of the DM falls below the escape velocity of the stellar object $v_{\rm esc}$ after the first scattering. For most of the relevant parameter space, the majority of DM capture occurs after the first scattering.

A key distinction arises between capture mediated by heavy and light particles. When the mediator is heavy compared to the momentum transfer, the fractional energy loss per collision is uniformly distributed, leading to a trivial form for the capture probability. In contrast, when the mediator is light, this distribution becomes momentum-dependent, modifying the capture probability. 
For instance, the single-scattering capture probability for DM interacting with nucleons through a heavy mediator is given by \cite{Dasgupta_2019}
\begin{eqnarray}
    g_1 (u) &=&\int_{0}^{1}\,dz\,\Theta \left(v_{\rm esc} - v_{\rm f}\right)\\
            &=&\dfrac{1}\beta\left(\beta - \dfrac{u^2}{u^2+v_{\rm esc}^2}\right)\,\Theta\left(\beta - \dfrac{u^2}{u^2+v_{\rm esc}^2}\right)\,,
            \label{eq:g1heavy}
\end{eqnarray}
where $0 \leq z \leq 1$ is the fractional loss in kinetic energy of the DM particle, $v_f = \sqrt{(1-z\,\beta)(u^2+v_{\rm esc}^2)}$ is the final reduced velocity after the first scattering and $\beta = 4\,m_\chi\,m_N / (m_\chi+m_N)$ with $m_N$ being the target nucleon mass.

When the mediator mass is light compared to the typical momentum transfer, the above expression is modified as \cite{Dasgupta_2020}
\begin{eqnarray}
    g_1 (u) &=&\int_{0}^{1}\,dz\,\Theta \left(v_{\rm esc} - v_{\rm f}\right)\,s(z)\,,
\end{eqnarray}
where $s(z) = \dfrac{m_\phi^2(\mu^2v_{\rm rel}^2+m_\phi^2)}{(\mu^2v_{\rm rel}^2\,z+m_\phi^2)^2}$ is a modulating function that encapsulates the effect of the light mediator with $\mu$ being the reduced mass and $v_{\rm rel}$ is the relative velocity between the incoming DM and the target. For heavy mediators, $s(z) \to 1$.
Upon integration, we obtain
\begin{eqnarray}
   g_1(u) &=&\dfrac{m_\phi^2\left(1-\dfrac{u^2}{\beta\,(u^2+v_{\rm esc}^2)}\right)}{\left(m_\phi^2+\dfrac{4\mu^2u^2}{\beta\,c^2}\right)}\,\Theta \left(v_{\rm esc}\sqrt{\beta/(1-\beta)}-u\right)\,.
\end{eqnarray}
In the limit $m_\phi \to \infty$, Eq.\,(\ref{eq:g1heavy}) is recovered, as expected.

As discussed before, the maximum allowed DM-nucleon cross section is given by $\sigma_{\rm sat}$, while for DM self-interactions, the corresponding saturation cross-section is $\sigma_{\rm s}^{\rm sat}$.  When $\sigma_{\chi\chi}$ exceeds $\sigma_{\chi\chi}^{\rm sat}$, the change in the DM population due to self-capture grows linearly with time. For smaller cross sections, the growth is exponential. This behaviour is naturally encoded in our analysis through the critical timescale $t_{\rm crit}$, as discussed earlier.

\subsection{SN-cooling with DM}
The population of DM particles inside the core of the SN can significantly modify its cooling via its emission of $A^\prime$. The trapped DM acts as an additional scattering target for $A'$ particles, thereby changing their free-streaming length.  The dominant scattering process is Thomson scattering between the DM and dark photons, $\chi\, A^\prime \rightarrow \chi\, A^\prime$. In contrast, bremsstrahlung processes such as $\chi\,\chi \rightarrow \chi\,\chi\, A^\prime$ are not effective in producing a significant population of $A'$ particles capable of affecting the cooling rate. This is because the rate of bremsstrahlung scales as the square of the captured DM density, whereas the density of captured DM inside the neutron star is relatively low. 
The Thomson scattering rate, on the other hand, scales as the captured DM density, making it the dominant process. 
Hence, $A'$ particles are much more likely to scatter off trapped DM than to be produced through DM--DM interactions.

In the absence of DM, the free streaming length of dark photons is $\lambda_p\equiv 1/(n_p\, \sigma_{p A'})$, where  $\sigma_{p A'} \propto \epsilon^2$ and cannot be too large because of upper limits on $\epsilon$ from existing terrestrial experiments and astrophysical sources. In the presence of DM, the free streaming length is modified to $\lambda_{\rm eff}$ as discussed in Eq.\,(\ref{eq:Lambdaeff}). Since there are no strong constraints on $\sigma_{\chi A'}$, this cross-section can be relatively large, compensating for the smallness of $n_\chi$ and potentially dominating the opacity of the medium. 
As a result, the effective free-streaming length of $A'$ within the SN core can be significantly shortened compared to the DM-free case.

If $\lambda_{\rm eff} < R_{\rm core}$, the dark photons become trapped and form a dark sphere inside the star, with a radius $r_{A'}$ given by
\begin{eqnarray}
    \int_{r_{A^\prime}}^{\infty}\left(n_\chi (r)\,\sigma_{\chi A^\prime} + n_p (r)\,\sigma_{p A^\prime}\right)\, dr = \dfrac{2}{3}\,.
\end{eqnarray}
In this regime, radiation of $A'$ occurs from the surface of this inner sphere rather than from the SN core surface itself. 
The radial distribution of the captured DM number density is given by Eq.\,(\ref{eq:DMdist}), while $\sigma_{\chi A^\prime}$ can be evaluated in our model to be $\sim \alpha^{\prime\,2} / m_\chi^2$. With this information, $r_{A'}$ can be obtained numerically from the above relation.

For very low DM densities inside the star, the resulting modification to the cooling rate is negligible, and the scenario effectively reduces to the standard case without DM. With increasing density, a significant $A^\prime$ sphere is formed with $r_{A^\prime} < R_{\rm core}$. The gauge bosons inside this sphere are trapped and cannot escape from the SN core, thereby suppressing energy loss. Only dark photons produced in the outer annular shell between  $R_{\rm core}$ and $r_{A'}$ contribute to the luminosity, which is then given by
\begin{eqnarray}
    L_{A^\prime} = \dfrac{4}{3}\pi \,\left(R_{\rm core}^3 -\,r_{A^\prime}^3\right)\,n_p^2\,T_c\, \sigma_{p A^\prime} \,e^{-m_{A^\prime}/T_c}\,e^{-\Gamma_{\rm decay}\,R_{\rm core}}\,.
\end{eqnarray}

At even higher DM densities, the $A^\prime$ sphere can extend outside the SN core. In this case, cooling happens due to surface emission and is governed by the Stefan-Boltzmann law. The luminosity is 
\begin{eqnarray}
    L_{A^\prime} = 4\,\pi\,\sigma_B \,\eta\, r_{A^\prime}^2\,T^4 = 4\,\pi\,\sigma_B \, \eta \, r_{A^\prime}^2\,T_c^4\,\left(\dfrac{r_{A^\prime}}{R_{\rm core}}\right)^{-20/3}\,e^{-m_{A^\prime}/T_c}\,,
    \end{eqnarray}
where $\eta$ is a suppression factor that takes into account the fact that, due to large DM densities, some of the $A^\prime$ particles are trapped inside the core and do not take part in surface emission. Therefore, $\eta$ can be interpreted as the fractional opacity and is given by $n_p\,\sigma_{p A^\prime}/(n_p\,\sigma_{p A^\prime} + n_\chi\,\sigma_{\chi A^\prime})$.

\begin{figure}[!t]
    \centering
    \subfigure[In absence of DM the total luminosity has contributions from both the surface and volume of the star depending on the free-streaming length $\lambda_{p}$. Volume emission dominates for small $\epsilon$ while for larger $\epsilon$ the emission is mostly from the surface. The surface luminosity follows the usual Stefan-Boltzmann law of radiation.]{\includegraphics[width=0.47\textwidth]{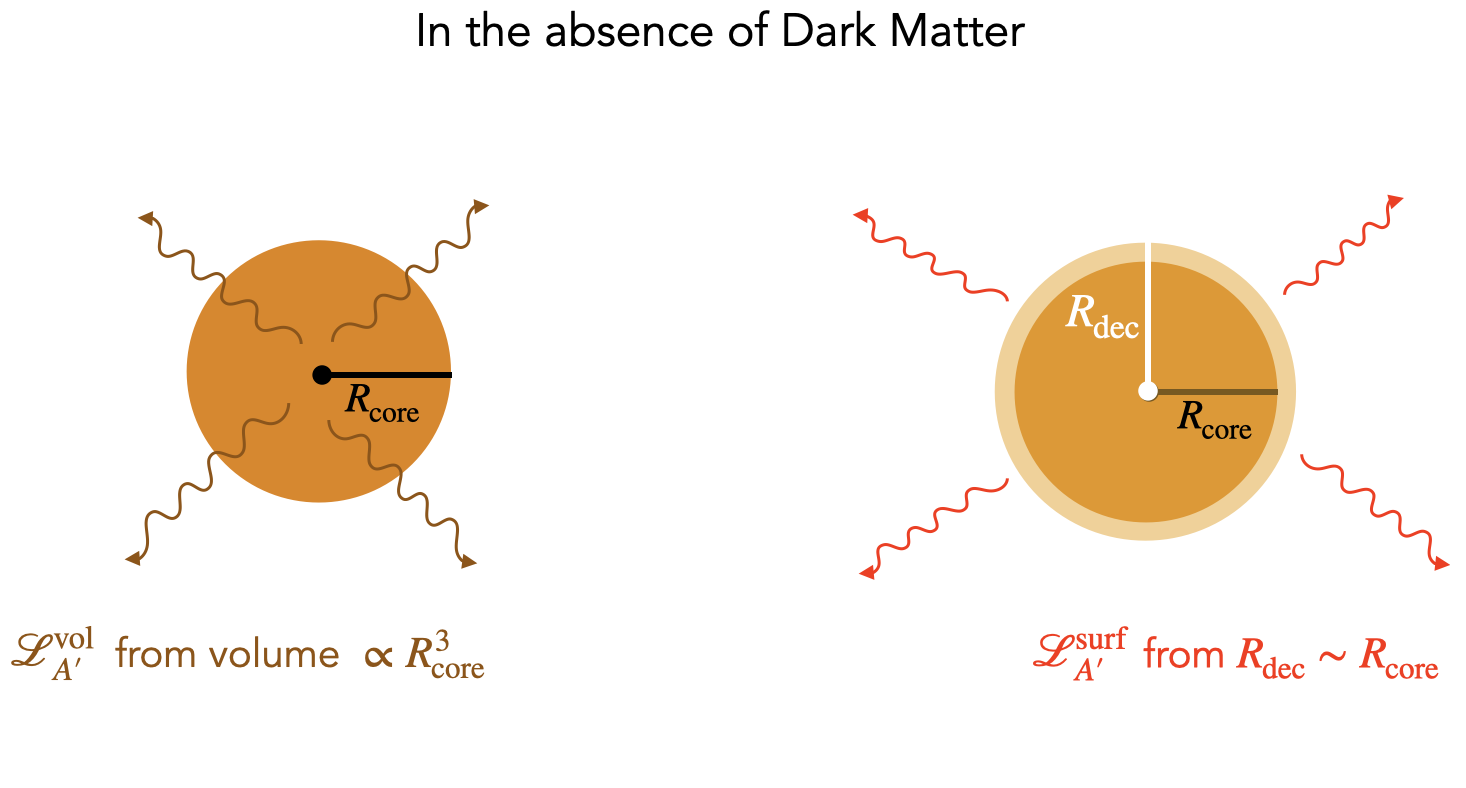}}
    \hfill
    \subfigure[With DM, the situation on the left changes. Depending on whether $\lambda_{p}$ is greater or less than $\lambda_{\chi}$, emission takes place either via the two modes as shown above. When $r_{A^\prime} > R_{\rm core}$, the emission takes place solely from the surface of this sphere following the Stefan-Boltzmann law. For $r_{A^\prime} < R_{\rm core}$, emission from the annular volume dominates.]{
        \includegraphics[width=0.47\textwidth]{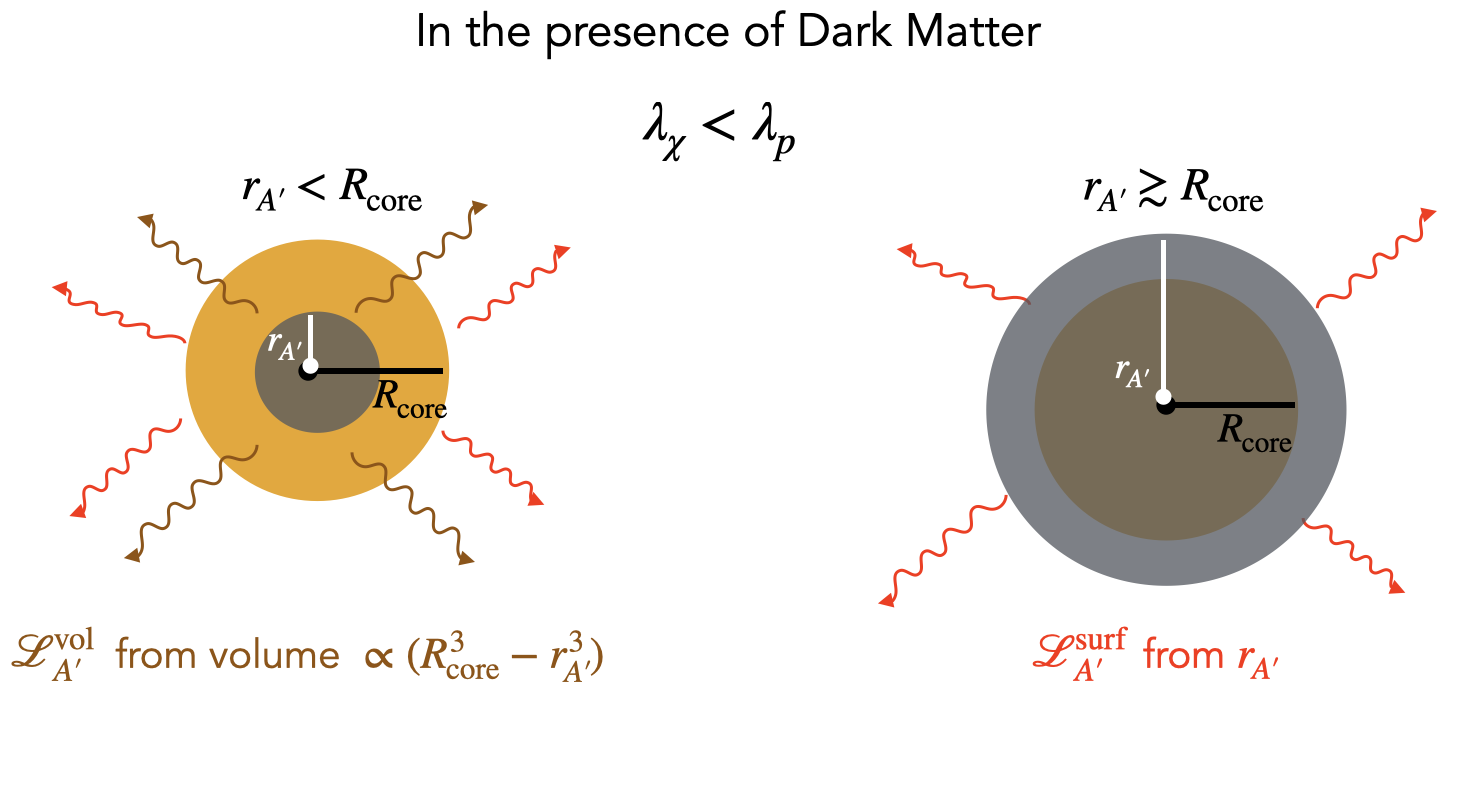}
    }
    \caption{Cartoon depicting how inclusion of DM changes the original cooling mechanism via $A^\prime$ emission in a typical neutron star. }
    \label{fig:cartoon}
\end{figure}

A schematic illustration of these regimes is shown in Fig.\,(\ref{fig:cartoon}). 
The model parameters are constrained by comparing the theoretically computed luminosity $L_{A'}$ with the observational upper limit of 
$3\times 10^{52}\,{\rm erg\,s^{-1}}$. The appropriate expression for $L_{A'}$ is used depending on whether $r_{A'}$ lies inside or outside the stellar core.

\section{Results}
\label{sec:4}
Before presenting our results, we stress that the precise numerical values of the exclusion regions depend on the simplified SN model adopted in this work. In particular, the inclusion of plasma effects and realistic radial profiles is expected to shift the boundaries quantitatively. Nevertheless, the qualitative features we identify—such as the emergence of a dark photosphere and the reopening of parameter space due to enhanced opacity—are robust consequences of the presence of an additional scattering channel and are therefore expected to persist in more detailed treatments.
\subsection{Annihilating DM}
\begin{figure}[!t]
\includegraphics[width=0.45\linewidth]{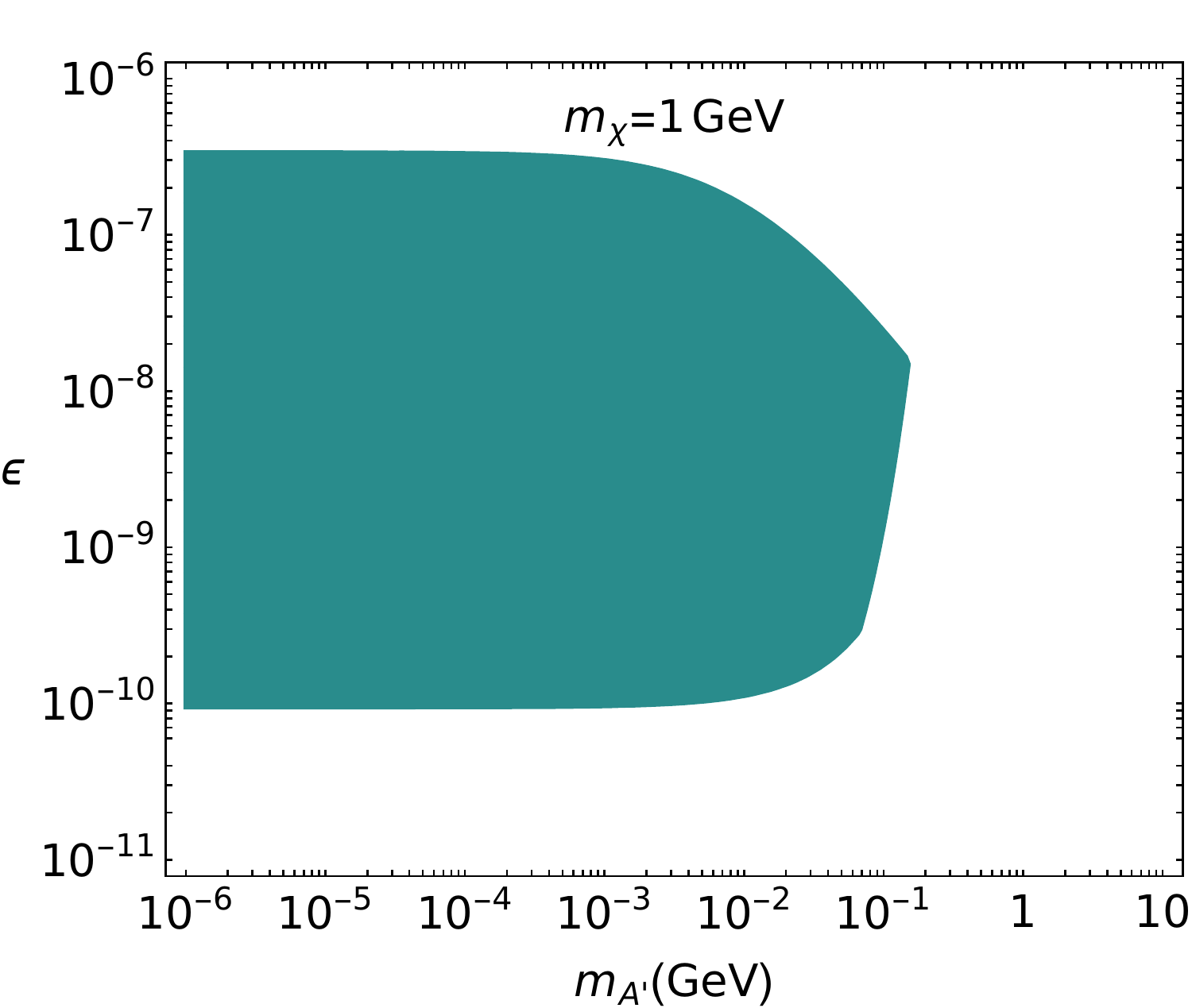}
    \caption{SN Cooling constraint for annihilating DM. The cross-section $\sigma_{\chi \chi} = 10^{-30}\,{\rm cm}^2$, and the coupling $\alpha'=0.03$. The constraint is similar in nature to the scenario with no captured dark matter. The upper boundary of the teal shaded portion denotes constraints from the surface emission, while the lower boundary is due to the exclusion from volume emission.}
     \label{fig:scananndm}
 \end{figure}

In this section, we discuss the impact of annihilating DM accumulated inside the SN core. As discussed in sec~\ref{sec:anndm}, the total number of captured DM particles saturates to a constant value at late times of the order of the age of the star. From a typical benchmark in this scenario, the final number of captured DM is found to be $\lesssim 10^{30}$. Together with the core temperature, this determines the spatial distribution of the DM number density inside the star. 
The corresponding number density, in turn, governs the free-streaming length of the dark photon. For the presence of DM to have a noticeable impact on the cooling mechanism, its contribution to the opacity must dominate over that of baryonic matter, i.e.,
$n_\chi\,\sigma_{\chi A'} \gg n_p\,\sigma_{pA'}$.  If the SN core is not able to accumulate enough DM, this inequality is not satisfied, and the modification to the cooling rate remains negligible.
We find that this is precisely the case with annihilating DM: a significant part of the captured DM is lost via its annihilation to particles that escape the SN core, leading to its decreased number density. The resulting exclusion region in the $(m_{A'},\,\epsilon)$ plane, shown in Fig.~(\ref{fig:scananndm}), closely resembles that obtained in the absence of DM capture.
We have checked that this conclusion remains robust for different combinations of model parameters as well. For example, increasing the DM mass reduces the capture rate in compact objects, resulting in a smaller number of captured DM particles and, therefore, a diminished impact on the cooling process. Conversely, lighter DM particles are captured more efficiently, resulting in a higher number density. 
However, a smaller $m_\chi$ simultaneously enhances the annihilation rate, which depletes the captured population more rapidly. 
As a result, the net effect remains largely unchanged. 
Explicit calculations with $m_\chi = 0.1\,{\rm GeV}$ confirm that the exclusion contour retains the same shape as in Fig.~(\ref{fig:scananndm}).

Varying the self-interaction cross-section also fails to improve the situation. As illustrated in Fig.~(\ref{fig:Nchi_t_ann}) reducing $\sigma_{\chi\chi}$ lowers the number of accreted DM particles. While increasing $\sigma_{\chi\chi}$ can enhance self-capture initially, this enhancement saturates once $\sigma_{\chi\chi}$ reaches the geometric limit. Beyond this point, no further increase in the total number of captured particles occurs. Even for the largest permissible self-interaction cross sections, of order $\sim 10^{-15}\,{\rm cm^2}$, we observe no significant rise in the DM population inside the star.


\subsection{Asymmetric DM}
The dynamics change considerably in the case of asymmetric DM. Unlike annihilating DM, which reaches a steady-state abundance due to self-annihilation, the number of asymmetric DM particles inside the SN core continues to grow with time, as there are no anti-DM particles available for annihilation.
As seen in Fig.~(\ref{fig:Nchi_t_ass}), the captured DM population increases monotonically and never saturates. Consequently, the accumulated DM can reach a significantly larger number, potentially leading to a non-trivial modification of the dark photon free-streaming length and hence of the stellar cooling rate. 

\begin{figure}[!t]
\includegraphics[width=0.45\linewidth]{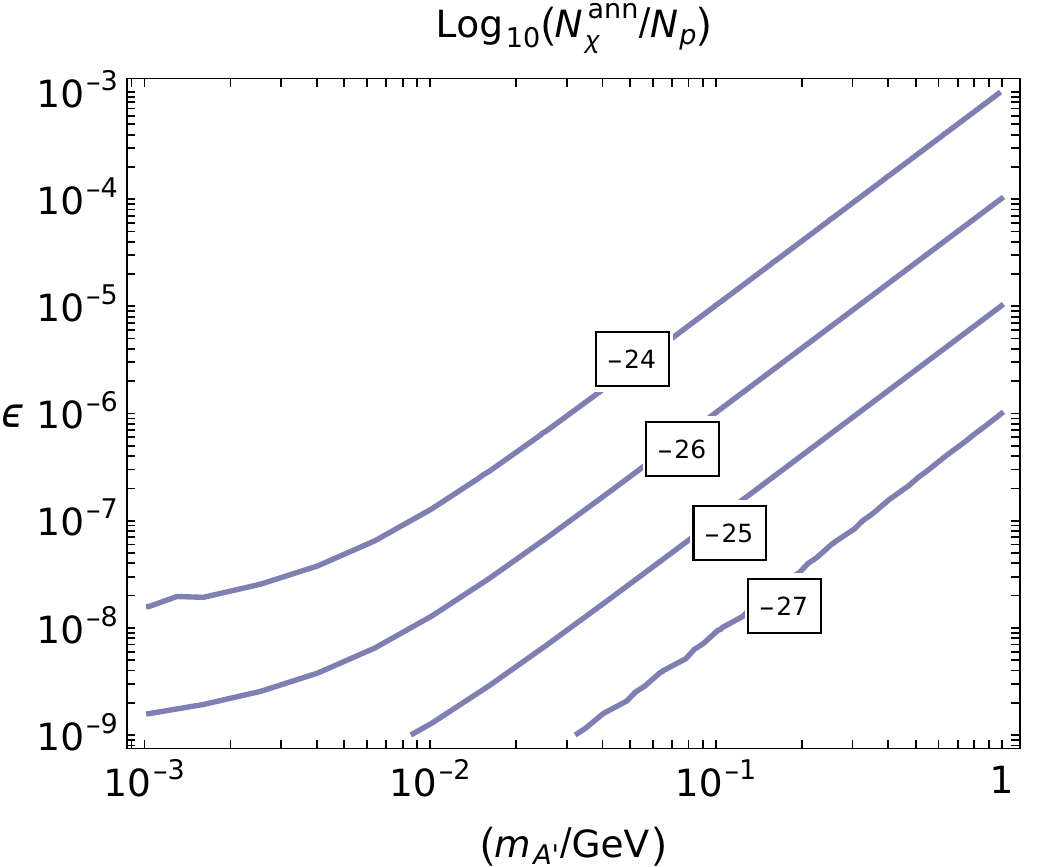}
~~\includegraphics[width=0.45\linewidth]{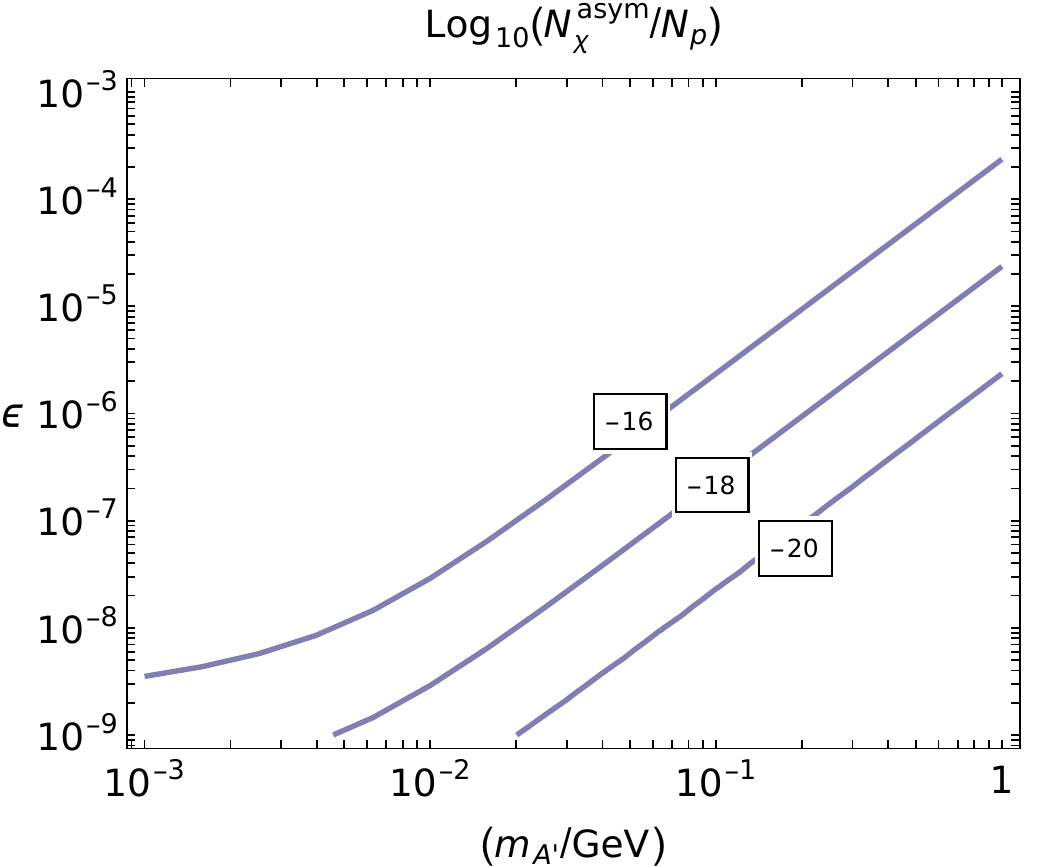}
    \caption{The ratio of the final number of captured DM particles with the total proton number, $N_\chi/N_p$ in the $\epsilon-m_{A'}$ plane for $m_\chi =100\,{\rm GeV}$. Left panel: Annihilating DM. Right panel: Asymmetric DM. As expected, annihilating dark matter is much less abundant inside the star as compared to its non-annihilating counterpart.}
     \label{fig:annVsasym}
 \end{figure}

To quantify this enhancement, we compare the total amount of captured DM in the annihilation and asymmetric case by plotting contours of the ratio of captured DM to the total number of nucleon targets in the $\epsilon-m_{A^\prime}$ plane in Fig.~(\ref{fig:annVsasym}). The results clearly demonstrate that the DM population within the stellar lifetime is substantially higher in the asymmetric scenario compared to the annihilating one.

\begin{figure}[!t]
\includegraphics[width=0.3\linewidth]{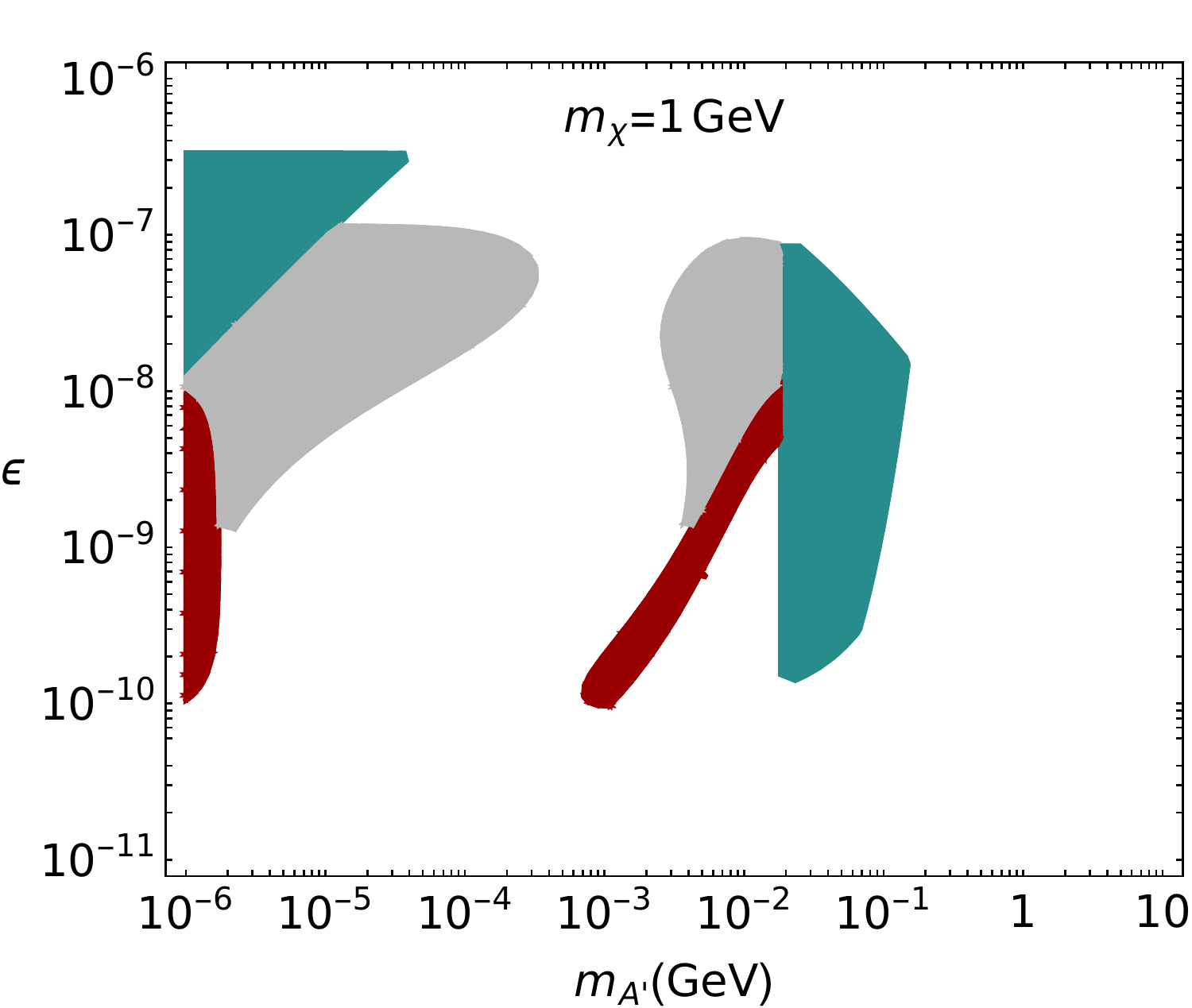}~~\includegraphics[width=0.3\linewidth]{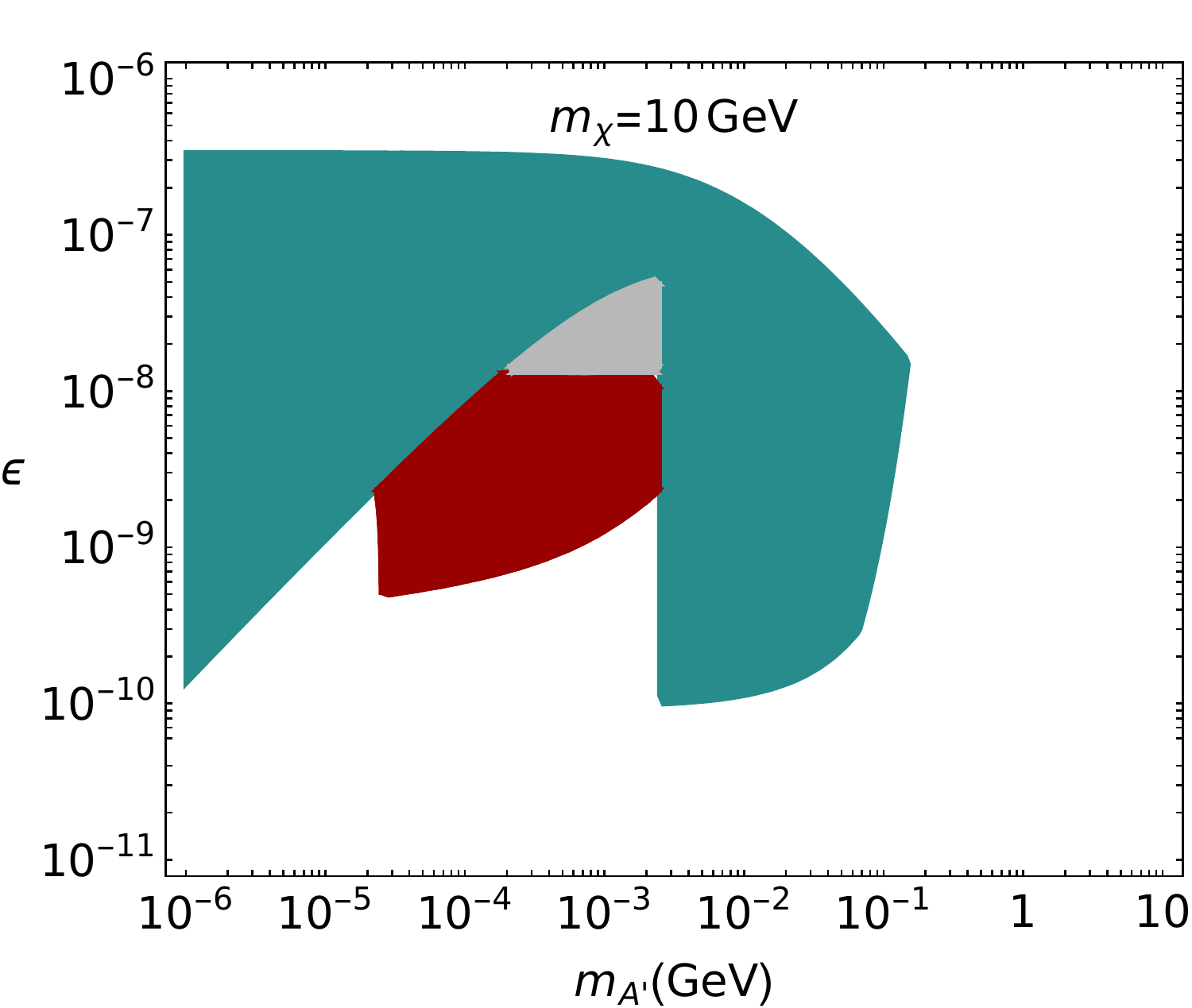}~~\includegraphics[width=0.3\linewidth]{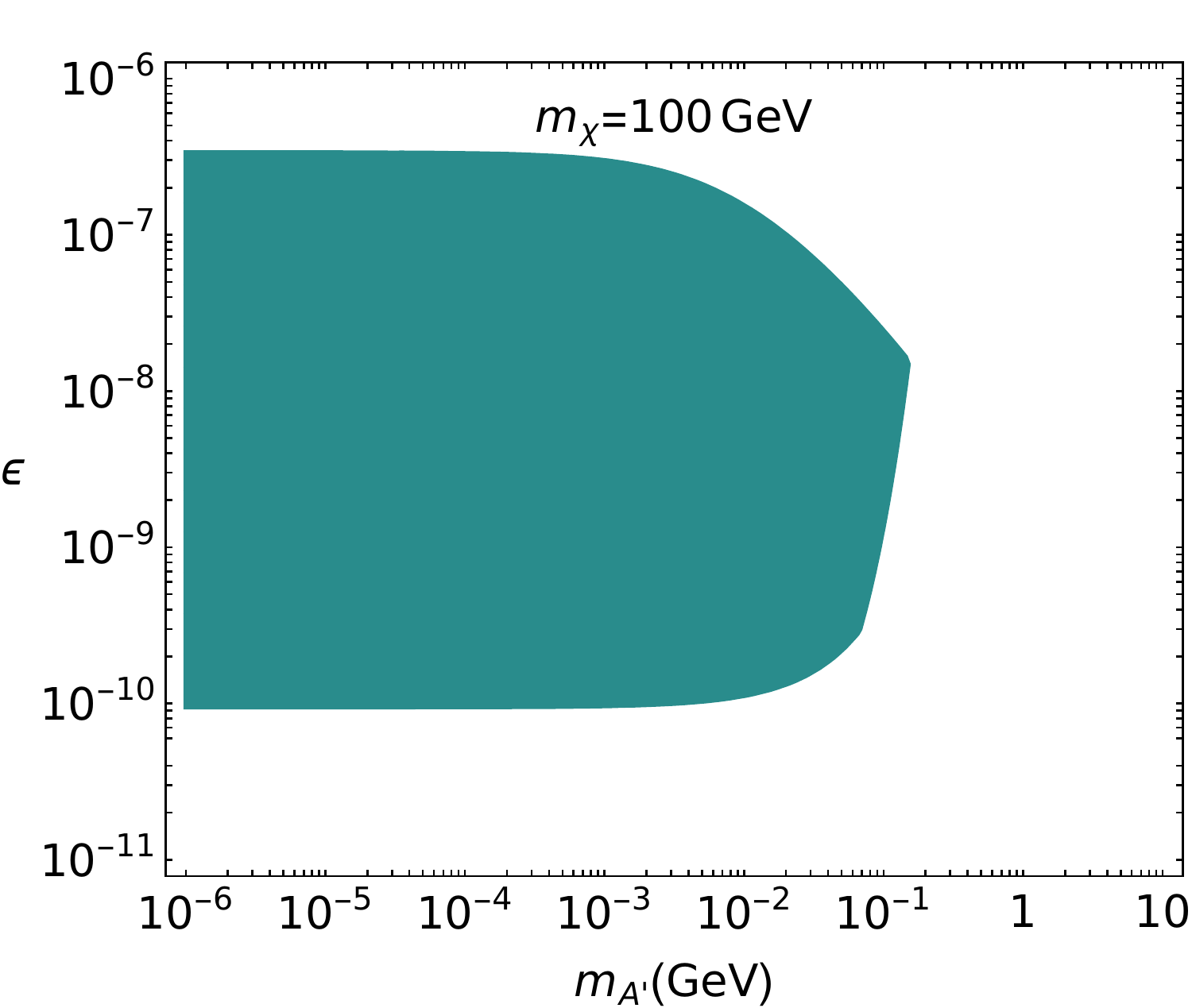}
    \caption{SN Cooling constraint for asymmetric DM. The cross-section $\sigma_{\chi \chi} = 10^{-30}\,{\rm cm}^2$, and the dark photon coupling $\alpha^\prime = 0.03$. Left panel: $m_\chi = 1\,{\rm GeV}$. Middle panel: $m_\chi = 10\,{\rm GeV}$. Right panel: $m_\chi = 100\,{\rm GeV}$. We find parts of the parameter space opening up as we decrease the mass of the asymmetric dark matter for reasons mentioned in the text. The \textcolor{teal}{\bf{teal-shaded}} region is ruled out when dark matter accumulation is not enough to produce any sizable improvement over the no-DM scenario. The \textcolor{purple}{\bf{maroon-shaded}} region is ruled out when luminosity due to emission from the annular volume between the surface of the $A^\prime$ sphere and the surface of the core exceeds the observed luminosity. Finally, the \textcolor{darkgray}{\bf{grey-shaded}} region is ruled out when the decoupling radius is larger than the core radius, and the surface-only emission supersedes the observed luminosity.}
     \label{fig:finscanasym}
 \end{figure}

This large accumulation of DM particles opens up previously ruled-out parts of the parameter space. We scan the $\epsilon-m_{A^\prime}$ plane and highlight in Fig.~(\ref{fig:finscanasym}) the regions where the theoretically predicted luminosity exceeds the observed value. For DM masses $\sim 100$ GeV, we find that the shape of the excluded (teal) region remains unchanged compared to the annihilating DM scenario. This is expected because heavier DM implies less effective capture, leading to poor accumulation of DM particles within the star. Therefore, due to this low density of DM, the situation is identical to the case with no or little DM accretion.

The picture changes when DM mass is reduced $(m_\chi \sim 10\,{\rm GeV})$. As shown in the middle panel of Fig.~(\ref {fig:finscanasym}), a new triangular, wedge-shaped region emerges in the parameter space (teal region). Within this region, dark photons that previously escaped freely from the neutron star become trapped by scatterings with the dense DM population, thereby suppressing the luminosity. For a fixed DM mass and kinetic-mixing parameter $\epsilon$, this transition occurs when $m_{A'} \lesssim 10^{-3}\,{\rm GeV}$. As $m_{A'}$ decreases, the capture rate initially increases because the scattering cross section $\sigma_{\chi p}$ grows for lighter mediators, allowing greater DM buildup and thus expanding the allowed region of parameter space consistent with observations. With further reduction in $m_{A'}$, the cross section hits the saturation limit and cannot grow further. Beyond this, the capture rate begins to fall since the probability of capture $g_1(u)$ decreases with decreasing mediator mass. This reduces the overall capture efficiency. Consequently, the previously opened parameter space closes again once $m_{A'}$ falls below a certain threshold. 
This turnover behaviour, absent in earlier studies, arises naturally here due to our more accurate treatment of the capture process in the light-mediator regime.

However, not all regions of this newly opened triangular wedge are available. As depicted schematically in Fig.~(\ref{fig:cartoon}), the total luminosity now receives contribution from emissions from the annular volume and is $\propto \left(R_{\rm core}^3-r_{A^\prime}^3\right)$ as long as $r_{A^\prime} < R_{\rm core}$. As the DM mass decreases, its number density increases, causing more $A'$ particles to scatter within the star, thereby enlarging $r_{A'}$. This leads to increased emission from the annular shell, and portions of the previously allowed parameter space (triangular, wedge-like region) become excluded. This region appears as the maroon-shaded area in Fig.~(\ref{fig:finscanasym}). 
For still higher $\epsilon$ values for a given $m_{A'}$, the decoupling radius $r_{A^\prime}$ increases further and at some point becomes larger than $R_{\rm core}$. Volume emission now becomes negligible, and surface emission starts dominating, resulting in a complementary grey exclusion region. Larger $\epsilon$ enhances $\sigma_{\chi p}$, which boosts the DM capture rate and number density, ultimately increasing the luminosity beyond observational limits and thereby excluding these parameters.

At even smaller DM masses $(m_\chi \sim 1\,{\rm GeV})$, the excluded parameter space separates into two disconnected regions, as seen from the leftmost plot of Fig.~\ref {fig:finscanasym}. Between these, a central white space appears where the number density of DM is high enough to trap the $A^\prime$ particles inside the star. In this region, both surface and volume emissions are suppressed, and $A'$ cannot contribute to the cooling. As $m_{A'}$ decreases further, the constraints tighten again, ruling out parts of the parameter space. 

In summary, a point in the parameter space is ruled out if any one of the following mutually exclusive conditions is satisfied:
\begin{enumerate}
    \item the DM number density is too low to alter the cooling rate appreciably;
    \item the total luminosity from $A'$ emission—arising from both the annular shell and the surface of the $A'$ sphere exceeds the observed luminosity; or
    \item the surface luminosity from an $A'$ sphere extending beyond the neutron star radius exceeds the observational limit.
\end{enumerate}
These three criteria correspond to the distinct colored regions shown in Fig.~(\ref{fig:finscanasym}).

All of the results discussed above assume a fixed dark-sector coupling $\alpha'$. 
Decreasing $\alpha'$ for a given DM mass, mediator mass, and kinetic mixing parameter $\epsilon$ suppresses $\sigma_{\chi A'}$. A smaller $\sigma_{\chi A'}$ increases $\lambda_{\chi}$, making free streaming more efficient and reducing the effect of dark matter. Thus, for small enough $\alpha^\prime$, we expect no difference from the scenario with no DM. 
Conversely, increasing $\alpha'$ enhances the interaction strength, shortens $\lambda_{\chi}$, and generally allows a larger portion of the parameter space to be reopened.

Note that the toy model considered in this work is treated in a phenomenological manner, with parameters chosen to highlight the impact of capture on SN cooling. The dark matter relic abundance, as well as constraints from direct and indirect detection, are determined by the structure of the complete underlying model. A kinetically mixed dark photon is usually subject to a wide range of laboratory constraints across several orders of magnitude in mass. These bounds arise from complementary experimental approaches, including $e^+e^-$ colliders, fixed-target and beam-dump experiments, as well as precision measurements, collectively probing the $(m_{A'}, \epsilon)$ parameter space. In the mass range $m_{A'} \sim \mathcal{O}(10\,{\rm MeV} - 10\,{\rm GeV})$, $e^+e^-$ collider experiments provide some of the most stringent constraints. In particular, monophoton searches at BaBar, based on the process $e^+e^- \to \gamma A'$ with $A'$ decaying invisibly, exclude kinetic mixing values $\epsilon \gtrsim 10^{-3} - 10^{-4}$ over a broad range of mediator masses~\cite{BaBar:2017tiz}.

Complementary searches at BESIII further probe the $\mathcal{O}(1\,{\rm GeV})$ regime, placing additional bounds on visibly and invisibly decaying dark photons~\cite{BESIII:2017fwv}. At higher masses, collider experiments such as LHCb constrain dark photons through dilepton resonance searches, excluding comparable values of $\epsilon$ in the GeV-scale region~\cite{LHCb:2017trq}. For $m_{A'} \lesssim \mathcal{O}(1\,{\rm GeV})$, fixed-target experiments provide leading sensitivity, particularly when the dark photon decays invisibly into $\chi \bar{\chi}$. The NA64 experiment, using a missing-energy technique in electron scattering, currently probes kinetic mixing values down to $\epsilon \sim 10^{-4}$ for $m_{A'} \sim \mathcal{O}(1 - 100\,{\rm MeV})$, excluding significant portions of parameter space relevant for light dark matter~\cite{NA64:2019auh}. At even smaller couplings, beam-dump experiments such as SLAC E137, E141, and Orsay constrain long-lived dark photons through their decay signatures, reaching sensitivities of $\epsilon \lesssim 10^{-7}$ in favorable regions of parameter space~\cite{Bauer:2018onh}.

In the lower mass regime $m_{A'} \sim \mathcal{O}({\rm keV-MeV})$, the sensitivity of high-energy experiments diminishes, and constraints are instead dominated by precision observables and low-energy probes. Measurements of the anomalous magnetic moments of the electron and muon provide robust constraints on light vector mediators, excluding $\epsilon \gtrsim 10^{-2} - 10^{-3}$ over a wide sub-MeV mass range~\cite{Pospelov:2008zw}. In addition, precision QED tests, including atomic spectroscopy and searches for deviations from Coulomb's law, constrain new light kinetically mixed gauge bosons coupled to electric charge~\cite{Jaeckel:2010ni}. At the lowest masses, $m_{A'} \sim \mathcal{O}({\rm keV})$, emerging low-threshold direct detection experiments sensitive to electron recoils, such as SENSEI and SuperCDMS, have begun to probe dark sector interactions mediated by very light dark photons, providing complementary sensitivity in scenarios with light DM~\cite{Essig:2015cda}.

In the present analysis, we do not attempt to identify a fully viable cosmological realization of such DM models across the entire parameter space considered. Instead, our results demonstrate the fact that, irrespective of these details, the presence of a pre-existing DM population inside the progenitor star can qualitatively alter the SN cooling constraints on dark photons. 
\section{Discussion and Conclusion}
\label{sec:5}
The results presented in this work demonstrate that the presence of DM captured inside supernova progenitors can induce non-trivial modifications to the cooling constraints for dark photons. In the absence of DM, the conventional SN1987A cooling argument yields the well-known exclusion regions in the $(m_{A'}, \epsilon)$ plane, determined by the competition between volume and surface emission of $A'$ from the proto-neutron star core. These limits, however, implicitly assume that nucleons are the sole scattering targets inside the stellar core, and no DM has been captured. 

The fact that the progenitor star exists inside the DM halo implies that DM can get captured in the stellar core due to scattering on nucleons. Our study shows that the standard cooling argument is altered once the contribution of captured DM is taken into account. The trapped DM provides an additional scattering channel for dark photons, altering their free-streaming length and, consequently, the cooling luminosity.  The impact depends sensitively on the number density of captured DM, which in turn depends on its mass, interaction cross section with nucleons, and self-interactions, and the nature of the DM—whether annihilating or asymmetric.
\MS{A key assumption of the present analysis is that the DM population captured during the progenitor phase remains gravitationally bound during the collapse. While this approximation is sufficient for isolating the effect of DM induced opacity, a fully self-consistent treatment of DM transport during core collapse remains an important direction for future work.}

For annihilating DM, the interplay between capture and annihilation restricts the total number of DM particles that can accumulate inside the star. Our numerical results indicate that the resulting equilibrium number density is generally insufficient to modify the $A^\prime$ free-streaming length substantially. As a result, the supernova cooling bounds remain essentially unchanged, and the canonical exclusion regions obtained from SN1987A observations continue to hold. 
This conclusion remains robust across a wide range of model parameters, including variations in DM mass and self-interaction cross section.

In contrast, asymmetric (non-annihilating) DM can accumulate continuously over the lifetime of the progenitor. In this case, the number density of captured DM can grow sufficiently large to reduce the free-streaming length of $A'$, leading to the formation of a ``dark photosphere'' within or around the SN core. This alters the emission geometry, suppressing the luminosity and effectively reopening parts of the $(m_{A'}, \epsilon)$ parameter space that would otherwise be excluded in the absence of DM. The size of this effect is strongly dependent on DM properties, including mass, self-interaction cross section, and the strength of the dark photon–DM coupling.

Our analysis underscores the following crucial points:
\begin{itemize}
\item The inclusion of DM capture in the progenitor core introduces a new, physically motivated uncertainty in the SN cooling bounds for dark photons.
\item The effect is negligible for annihilating DM due to annihilation losses, but can be substantial for non-annihilating DM.
\item The interplay between the DM number density and the dark photon scattering cross section with DM controls the modification of the exclusion region.
\item These findings highlight the broader importance of incorporating astrophysical environments into dark sector phenomenology, particularly for scenarios involving feebly interacting particles coupled to DM.
\end{itemize}

In conclusion, our study provides a self-consistent analysis of the role of captured DM in modifying SN cooling constraints for dark photons. Using the dark photon as a representative and well-motivated example, we explicitly demonstrate the relevance of this effect. While annihilating DM does not significantly alter the standard bounds, asymmetric DM can reopen previously excluded regions in the dark photon parameter space. This work demonstrates the necessity of accounting for astrophysical DM populations in indirect searches for new light particles. For the dark matter mass considered in this work, they are mostly captured after the first scattering. For heavier dark matter particles, multiple scattering becomes important and should be accounted for when calculating capture rates. However, in such cases, we only expect mild improvements to the constraints derived here. Future studies incorporating more detailed SN simulations, plasma effects, and realistic radial profiles will allow for more precise and potentially stronger constraints, further clarifying the interplay between DM and feebly interacting particles in extreme astrophysical environments.

\section*{Acknowledgments}
We thank Sergio Palomarez Ruiz, Joachim Kopp and Ranjan Laha for useful discussions and suggestions. AG acknowledges the support of the Krea Research Grant, which facilitated an academic visit to IIT Bombay, where a substantial portion of the work was conducted. This support was instrumental in enabling the timely completion of the project. AG is partly supported by grant PID2023-151418NB-I00, which is funded by MCIU/AEI/10.13039/501100011033/ FEDER, UE.
MS acknowledges support from the Early Career Research Grant by Anusandhan National Research Foundation (project number ANRF/ECRG/2024/000522/PMS). MS also acknowledges support from the IoE-funded Seed funding for Collaboration and Partnership Projects - Phase-IV SCPP grant (RD/0524-IOE00I0-012) by IIT Bombay.

\bibliographystyle{apsrev4-1}
\bibliography{main} 

@ARTICLE{spergel,
   author = {{Press}, W.~H. and {Spergel}, D.~N.},
    title = "{Capture by the sun of a galactic population of weakly interacting, massive particles}",
  journal = {apj},
     year = 1985,
    month = sep,
   volume = 296,
    pages = {679-684},
      doi = {10.1086/163485},
   adsurl = {http://adsabs.harvard.edu/abs/1985ApJ...296..679P}
}

@ARTICLE{gould1,
   author = {{Gould}, A.},
    title = "{Resonant enhancements in weakly interacting massive particle capture by the earth}",
  journal = {apj},
     year = 1987,
    month = oct,
   volume = 321,
    pages = {571-585},
      doi = {10.1086/165653},
   adsurl = {http://adsabs.harvard.edu/abs/1987ApJ...321..571G}
}

@ARTICLE{gould2,
   author = {{Gould}, A.},
    title = "{Direct and indirect capture of weakly interacting massive particles by the earth}",
  journal = {apj},
     year = 1988,
    month = may,
   volume = 328,
    pages = {919-939},
      doi = {10.1086/166347},
   adsurl = {http://adsabs.harvard.edu/abs/1988ApJ...328..919G}
}

@ARTICLE{gould3,
   author = {{Gould}, A.},
    title = "{Gravitational diffusion of solar system WIMPs}",
  journal = {apj},
     year = 1991,
    month = feb,
   volume = 368,
    pages = {610-615},
      doi = {10.1086/169726},
   adsurl = {http://adsabs.harvard.edu/abs/1991ApJ...368..610G}
}

@article{Silk:1985ax,
      author         = "Silk, Joseph and Olive, Keith A. and Srednicki, Mark",
      title          = "{The Photino, the Sun and High-Energy Neutrinos}",
      journal        = "Phys. Rev. Lett.",
      volume         = "55",
      year           = "1985",
      pages          = "257-259",
      doi            = "10.1103/PhysRevLett.55.257",
      note           = "[,283(1985)]",
      reportNumber   = "FERMILAB-PUB-85-062-A",
      SLACcitation   = "%%CITATION = PRLTA,55,257;%%"
}

@article{Krauss:1985aaa,
      author         = "Krauss, Lawrence M. and Srednicki, Mark and Wilczek,
                        Frank",
      title          = "{Solar System Constraints and Signatures for Dark Matter
                        Candidates}",
      journal        = "Phys. Rev.",
      volume         = "D33",
      year           = "1986",
      pages          = "2079-2083",
      doi            = "10.1103/PhysRevD.33.2079",
      reportNumber   = "YTP-85-19",
      SLACcitation   = "%%CITATION = PHRVA,D33,2079;%%"
}

@article{PhysRevLett.60.1797,
  title = {Axions from SN1987A},
  author = {Turner, Michael S.},
  journal = {Phys. Rev. Lett.},
  volume = {60},
  issue = {18},
  pages = {1797--1800},
  numpages = {0},
  year = {1988},
  month = {May},
  publisher = {American Physical Society},
  doi = {10.1103/PhysRevLett.60.1797},
  url = {https://link.aps.org/doi/10.1103/PhysRevLett.60.1797}
}

@article{1703.04043,
      author         = "Bramante, Joseph and Delgado, Antonio and Martin, Adam",
      title          = "{Multiscatter stellar capture of dark matter}",
      journal        = "Phys. Rev.",
      volume         = "D96",
      year           = "2017",
      number         = "6",
      pages          = "063002",
      doi            = "10.1103/PhysRevD.96.063002",
      eprint         = "1703.04043",
      archivePrefix  = "arXiv",
      primaryClass   = "hep-ph",
      SLACcitation   = "%%CITATION = ARXIV:1703.04043;%%"
}

@article{Dasgupta_2019,
   title={Dark matter capture in celestial objects: improved treatment of multiple scattering and updated constraints from white dwarfs},
   volume={2019},
   ISSN={1475-7516},
   url={http://dx.doi.org/10.1088/1475-7516/2019/08/018},
   DOI={10.1088/1475-7516/2019/08/018},
   number={08},
   journal={Journal of Cosmology and Astroparticle Physics},
   publisher={IOP Publishing},
   author={Dasgupta, Basudeb and Gupta, Aritra and Ray, Anupam},
   year={2019},
   month=aug, pages={018–018} }

@INPROCEEDINGS{1987ESOC...26..237A,
       author = {{Alekseev}, E.~N. and {Alekseeva}, L.~N. and {Krivosheina}, I.~V. and {Volchenko}, V.~I.},
        title = "{Detection of the Neutrino Signal from Supernova 1987A Using the INR Baksan Underground Scintillation Telescope}",
    booktitle = {European Southern Observatory Conference and Workshop Proceedings},
         year = 1987,
       editor = {{Danziger}, I.~J.},
       series = {European Southern Observatory Conference and Workshop Proceedings},
       volume = {26},
        month = jan,
        pages = {237},
       adsurl = {https://ui.adsabs.harvard.edu/abs/1987ESOC...26..237A}
}

@article{Petraki_2017,
   title={Radiative bound-state-formation cross-sections for dark matter interacting via a Yukawa potential},
   volume={2017},
   ISSN={1029-8479},
   url={http://dx.doi.org/10.1007/JHEP04(2017)077},
   DOI={10.1007/jhep04(2017)077},
   number={4},
   journal={Journal of High Energy Physics},
   publisher={Springer Science and Business Media LLC},
   author={Petraki, Kalliopi and Postma, Marieke and de Vries, Jordy},
   year={2017},
   month=apr }

@article{Petraki_2015,
   title={Dark-matter bound states from Feynman diagrams},
   volume={2015},
   ISSN={1029-8479},
   url={http://dx.doi.org/10.1007/JHEP06(2015)128},
   DOI={10.1007/jhep06(2015)128},
   number={6},
   journal={Journal of High Energy Physics},
   publisher={Springer Science and Business Media LLC},
   author={Petraki, Kalliopi and Postma, Marieke and Wiechers, Michael},
   year={2015},
   month=jun }

@article{Dasgupta_2020,
   title={Dark matter capture in celestial objects: light mediators, self-interactions, and complementarity with direct detection},
   volume={2020},
   ISSN={1475-7516},
   url={http://dx.doi.org/10.1088/1475-7516/2020/10/023},
   DOI={10.1088/1475-7516/2020/10/023},
   number={10},
   journal={Journal of Cosmology and Astroparticle Physics},
   publisher={IOP Publishing},
   author={Dasgupta, Basudeb and Gupta, Aritra and Ray, Anupam},
   year={2020},
   month=oct, pages={023–023} }

@article{Kazanas_2015,
   title={Supernova bounds on the dark photon using its electromagnetic decay},
   volume={890},
   ISSN={0550-3213},
   url={http://dx.doi.org/10.1016/j.nuclphysb.2014.11.009},
   DOI={10.1016/j.nuclphysb.2014.11.009},
   journal={Nuclear Physics B},
   publisher={Elsevier BV},
   author={Kazanas, Demos and Mohapatra, Rabindra N. and Nussinov, Shmuel and Teplitz, Vigdor L. and Zhang, Yongchao},
   year={2015},
   month=jan, pages={17–29} }

@misc{dent2012,
      title={Constraints on Light Hidden Sector Gauge Bosons from Supernova Cooling}, 
      author={James B. Dent and Francesc Ferrer and Lawrence M. Krauss},
      year={2012},
      eprint={1201.2683},
      archivePrefix={arXiv},
      primaryClass={astro-ph.CO},
      url={https://arxiv.org/abs/1201.2683}, 
}

@article{Hardy_2017,
   title={Stellar cooling bounds on new light particles: plasma mixing effects},
   volume={2017},
   ISSN={1029-8479},
   url={http://dx.doi.org/10.1007/JHEP02(2017)033},
   DOI={10.1007/jhep02(2017)033},
   number={2},
   journal={Journal of High Energy Physics},
   publisher={Springer Science and Business Media LLC},
   author={Hardy, Edward and Lasenby, Robert},
   year={2017},
   month=feb }

@article{Hardy:2024gwy,
    author = "Hardy, Edward and Sokolov, Anton and Stubbs, Henry",
    title = "{Supernova bounds on new scalars from resonant and soft emission}",
    eprint = "2410.17347",
    archivePrefix = "arXiv",
    primaryClass = "hep-ph",
    doi = "10.1007/JHEP04(2025)013",
    journal = "JHEP",
    volume = "04",
    pages = "013",
    year = "2025"
}

@ARTICLE{2025JCAP...01..061F,
       author = {{Fischer}, Tobias and {Camalich}, Jorge Martin and {Kochankovski}, Hristijan and {Tolos}, Laura},
        title = "{Hyperons during proto-neutron star deleptonization and the emission of dark flavoured particles}",
      journal = {\jcap},
         year = 2025,
        month = jan,
       volume = {2025},
       number = {1},
          eid = {061},
        pages = {061},
          doi = {10.1088/1475-7516/2025/01/061},
archivePrefix = {arXiv},
       eprint = {2408.01406},
 primaryClass = {astro-ph.HE},
       adsurl = {https://ui.adsabs.harvard.edu/abs/2025JCAP...01..061F}
}

@article{Kamiokande-II:1987idp,
    author = "Hirata, K. and others",
    editor = "Wali, K. C.",
    collaboration = "Kamiokande-II",
    title = "{Observation of a Neutrino Burst from the Supernova SN 1987a}",
    reportNumber = "UT-ICEPP-87-01, UPR-142E",
    doi = "10.1103/PhysRevLett.58.1490",
    journal = "Phys. Rev. Lett.",
    volume = "58",
    pages = "1490--1493",
    year = "1987"
}

@article{Bionta:1987qt,
    author = "Bionta, R. M. and others",
    title = "{Observation of a Neutrino Burst in Coincidence with Supernova SN 1987a in the Large Magellanic Cloud}",
    reportNumber = "UCI-NEUTRINO-87-10",
    doi = "10.1103/PhysRevLett.58.1494",
    journal = "Phys. Rev. Lett.",
    volume = "58",
    pages = "1494",
    year = "1987"
}

@article{Horiuchi:2018ofe,
    author = "Horiuchi, Shunsaku and Kneller, James P",
    title = "{What can be learned from a future supernova neutrino detection?}",
    eprint = "1709.01515",
    archivePrefix = "arXiv",
    primaryClass = "astro-ph.HE",
    doi = "10.1088/1361-6471/aaa90a",
    journal = "J. Phys. G",
    volume = "45",
    number = "4",
    pages = "043002",
    year = "2018"
}

@article{Volpe:2023met,
    author = "Volpe, M. Cristina",
    title = "{Neutrinos from dense environments: Flavor mechanisms, theoretical approaches, observations, and new directions}",
    eprint = "2301.11814",
    archivePrefix = "arXiv",
    primaryClass = "hep-ph",
    doi = "10.1103/RevModPhys.96.025004",
    journal = "Rev. Mod. Phys.",
    volume = "96",
    number = "2",
    pages = "025004",
    year = "2024"
}

@article{Sen:2024fxa,
    author = "Sen, Manibrata",
    title = "{Supernova Neutrinos: Flavour Conversion Mechanisms and New Physics Scenarios}",
    eprint = "2405.20432",
    archivePrefix = "arXiv",
    primaryClass = "hep-ph",
    doi = "10.3390/universe10060238",
    journal = "Universe",
    volume = "10",
    number = "6",
    pages = "238",
    year = "2024"
}

@article{Raffelt:1987yt,
    author = "Raffelt, Georg and Seckel, David",
    title = "{Bounds on Exotic Particle Interactions from SN 1987a}",
    reportNumber = "SCIPP-87/107",
    doi = "10.1103/PhysRevLett.60.1793",
    journal = "Phys. Rev. Lett.",
    volume = "60",
    pages = "1793",
    year = "1988"
}

@article{Ellis:1987pk,
    author = "Ellis, John R. and Olive, Keith A.",
    title = "{Constraints on Light Particles From Supernova Sn1987a}",
    reportNumber = "CERN-TH-4701/87, UMN-TH-605/87",
    doi = "10.1016/0370-2693(87)91710-2",
    journal = "Phys. Lett. B",
    volume = "193",
    pages = "525",
    year = "1987"
}

@article{Burrows:1990pk,
    author = "Burrows, Adam and Ressell, M. Ted and Turner, Michael S.",
    title = "{Axions and SN1987A: Axion trapping}",
    reportNumber = "FERMILAB-PUB-90-081-A, FERMILAB-PUB-90-081-A-REV",
    doi = "10.1103/PhysRevD.42.3297",
    journal = "Phys. Rev. D",
    volume = "42",
    pages = "3297--3309",
    year = "1990"
}

@book{Raffelt:1996wa,
    author = "Raffelt, G. G.",
    title = "{Stars as laboratories for fundamental physics}: {The astrophysics of neutrinos, axions, and other weakly interacting particles}",
    isbn = "978-0-226-70272-8",
    month = "5",
    year = "1996"
}

@article{Dolan:2017osp,
    author = "Dolan, Matthew J. and Ferber, Torben and Hearty, Christopher and Kahlhoefer, Felix and Schmidt-Hoberg, Kai",
    title = "{Revised constraints and Belle II sensitivity for visible and invisible axion-like particles}",
    eprint = "1709.00009",
    archivePrefix = "arXiv",
    primaryClass = "hep-ph",
    reportNumber = "DESY-17-127",
    doi = "10.1007/JHEP12(2017)094",
    journal = "JHEP",
    volume = "12",
    pages = "094",
    year = "2017",
    note = "[Erratum: JHEP 03, 190 (2021)]"
}

@article{Chang:2018rso,
    author = "Chang, Jae Hyeok and Essig, Rouven and McDermott, Samuel D.",
    title = "{Supernova 1987A Constraints on Sub-GeV Dark Sectors, Millicharged Particles, the QCD Axion, and an Axion-like Particle}",
    eprint = "1803.00993",
    archivePrefix = "arXiv",
    primaryClass = "hep-ph",
    reportNumber = "YITP-SB-18-01, FERMILAB-PUB-17-432-T",
    doi = "10.1007/JHEP09(2018)051",
    journal = "JHEP",
    volume = "09",
    pages = "051",
    year = "2018"
}

@article{Carenza:2019pxu,
    author = "Carenza, Pierluca and Fischer, Tobias and Giannotti, Maurizio and Guo, Gang and Mart{\'\i}nez-Pinedo, Gabriel and Mirizzi, Alessandro",
    title = "{Improved axion emissivity from a supernova via nucleon-nucleon bremsstrahlung}",
    eprint = "1906.11844",
    archivePrefix = "arXiv",
    primaryClass = "hep-ph",
    doi = "10.1088/1475-7516/2019/10/016",
    journal = "JCAP",
    volume = "10",
    number = "10",
    pages = "016",
    year = "2019",
    note = "[Erratum: JCAP 05, E01 (2020)]"
}

@article{Carenza:2020cis,
    author = "Carenza, Pierluca and Fore, Bryce and Giannotti, Maurizio and Mirizzi, Alessandro and Reddy, Sanjay",
    title = "{Enhanced Supernova Axion Emission and its Implications}",
    eprint = "2010.02943",
    archivePrefix = "arXiv",
    primaryClass = "hep-ph",
    reportNumber = "INT-PUB-20-039",
    doi = "10.1103/PhysRevLett.126.071102",
    journal = "Phys. Rev. Lett.",
    volume = "126",
    number = "7",
    pages = "071102",
    year = "2021"
}

@article{Dev:2020eam,
    author = "Dev, P. S. Bhupal and Mohapatra, Rabindra N. and Zhang, Yongchao",
    title = "{Revisiting supernova constraints on a light CP-even scalar}",
    eprint = "2005.00490",
    archivePrefix = "arXiv",
    primaryClass = "hep-ph",
    doi = "10.1088/1475-7516/2020/08/003",
    journal = "JCAP",
    volume = "08",
    pages = "003",
    year = "2020",
    note = "[Erratum: JCAP 11, E01 (2020)]"
}

@article{Lucente:2020whw,
    author = "Lucente, Giuseppe and Carenza, Pierluca and Fischer, Tobias and Giannotti, Maurizio and Mirizzi, Alessandro",
    title = "{Heavy axion-like particles and core-collapse supernovae: constraints and impact on the explosion mechanism}",
    eprint = "2008.04918",
    archivePrefix = "arXiv",
    primaryClass = "hep-ph",
    doi = "10.1088/1475-7516/2020/12/008",
    journal = "JCAP",
    volume = "12",
    pages = "008",
    year = "2020"
}

@article{Calore:2021klc,
    author = "Calore, Francesca and Carenza, Pierluca and Giannotti, Maurizio and Jaeckel, Joerg and Lucente, Giuseppe and Mirizzi, Alessandro",
    title = "{Supernova bounds on axionlike particles coupled with nucleons and electrons}",
    eprint = "2107.02186",
    archivePrefix = "arXiv",
    primaryClass = "hep-ph",
    doi = "10.1103/PhysRevD.104.043016",
    journal = "Phys. Rev. D",
    volume = "104",
    number = "4",
    pages = "043016",
    year = "2021"
}

@article{Caputo:2024oqc,
    author = "Caputo, Andrea and Raffelt, Georg",
    title = "{Astrophysical Axion Bounds: The 2024 Edition}",
    eprint = "2401.13728",
    archivePrefix = "arXiv",
    primaryClass = "hep-ph",
    reportNumber = "MPP-2024-13, CERN-TH-2024-013",
    doi = "10.22323/1.454.0041",
    journal = "PoS",
    volume = "COSMICWISPers",
    pages = "041",
    year = "2024"
}

@article{Dent:2012mx,
    author = "Dent, James B. and Ferrer, Francesc and Krauss, Lawrence M.",
    title = "{Constraints on Light Hidden Sector Gauge Bosons from Supernova Cooling}",
    eprint = "1201.2683",
    archivePrefix = "arXiv",
    primaryClass = "astro-ph.CO",
    month = "1",
    year = "2012"
}

@article{Kazanas:2014mca,
    author = "Kazanas, Demos and Mohapatra, Rabindra N. and Nussinov, Shmuel and Teplitz, Vigdor L. and Zhang, Yongchao",
    title = "{Supernova Bounds on the Dark Photon Using its Electromagnetic Decay}",
    eprint = "1410.0221",
    archivePrefix = "arXiv",
    primaryClass = "hep-ph",
    reportNumber = "UMD-PP--014-015",
    doi = "10.1016/j.nuclphysb.2014.11.009",
    journal = "Nucl. Phys. B",
    volume = "890",
    pages = "17--29",
    year = "2014"
}

@article{Rrapaj:2015wgs,
    author = "Rrapaj, Ermal and Reddy, Sanjay",
    title = "{Nucleon-nucleon bremsstrahlung of dark gauge bosons and revised supernova constraints}",
    eprint = "1511.09136",
    archivePrefix = "arXiv",
    primaryClass = "nucl-th",
    reportNumber = "INT-PUB-15-065",
    doi = "10.1103/PhysRevC.94.045805",
    journal = "Phys. Rev. C",
    volume = "94",
    number = "4",
    pages = "045805",
    year = "2016"
}

@article{Chang:2016ntp,
    author = "Chang, Jae Hyeok and Essig, Rouven and McDermott, Samuel D.",
    title = "{Revisiting Supernova 1987A Constraints on Dark Photons}",
    eprint = "1611.03864",
    archivePrefix = "arXiv",
    primaryClass = "hep-ph",
    reportNumber = "YITP-SB-16-44",
    doi = "10.1007/JHEP01(2017)107",
    journal = "JHEP",
    volume = "01",
    pages = "107",
    year = "2017"
}

@article{Mahoney:2017jqk,
    author = "Mahoney, Cameron and Leibovich, Adam K. and Zentner, Andrew R.",
    title = "{Updated Constraints on Self-Interacting Dark Matter from Supernova 1987A}",
    eprint = "1706.08871",
    archivePrefix = "arXiv",
    primaryClass = "hep-ph",
    doi = "10.1103/PhysRevD.96.043018",
    journal = "Phys. Rev. D",
    volume = "96",
    number = "4",
    pages = "043018",
    year = "2017"
}

@article{DeRocco:2019jti,
    author = "DeRocco, William and Graham, Peter W. and Kasen, Daniel and Marques-Tavares, Gustavo and Rajendran, Surjeet",
    title = "{Supernova signals of light dark matter}",
    eprint = "1905.09284",
    archivePrefix = "arXiv",
    primaryClass = "hep-ph",
    doi = "10.1103/PhysRevD.100.075018",
    journal = "Phys. Rev. D",
    volume = "100",
    number = "7",
    pages = "075018",
    year = "2019"
}

@article{Croon:2020lrf,
    author = "Croon, Djuna and Elor, Gilly and Leane, Rebecca K. and McDermott, Samuel D.",
    title = "{Supernova Muons: New Constraints on $Z$' Bosons, Axions and ALPs}",
    eprint = "2006.13942",
    archivePrefix = "arXiv",
    primaryClass = "hep-ph",
    reportNumber = "MIT-CTP/5214, FERMILAB-PUB-20-246-A-T",
    doi = "10.1007/JHEP01(2021)107",
    journal = "JHEP",
    volume = "01",
    pages = "107",
    year = "2021"
}

@article{Sung:2021swd,
    author = "Sung, Allan and Guo, Gang and Wu, Meng-Ru",
    title = "{Supernova Constraint on Self-Interacting Dark Sector Particles}",
    eprint = "2102.04601",
    archivePrefix = "arXiv",
    primaryClass = "hep-ph",
    doi = "10.1103/PhysRevD.103.103005",
    journal = "Phys. Rev. D",
    volume = "103",
    number = "10",
    pages = "103005",
    year = "2021"
}

@article{Caputo:2021eaa,
    author = "Caputo, Andrea and Millar, Alexander J. and O'Hare, Ciaran A. J. and Vitagliano, Edoardo",
    title = "{Dark photon limits: A handbook}",
    eprint = "2105.04565",
    archivePrefix = "arXiv",
    primaryClass = "hep-ph",
    reportNumber = "NORDITA-2021-036",
    doi = "10.1103/PhysRevD.104.095029",
    journal = "Phys. Rev. D",
    volume = "104",
    number = "9",
    pages = "095029",
    year = "2021"
}

@article{Caputo:2021rux,
    author = "Caputo, Andrea and Raffelt, Georg and Vitagliano, Edoardo",
    title = "{Muonic boson limits: Supernova redux}",
    eprint = "2109.03244",
    archivePrefix = "arXiv",
    primaryClass = "hep-ph",
    reportNumber = "MPP-2021-154",
    doi = "10.1103/PhysRevD.105.035022",
    journal = "Phys. Rev. D",
    volume = "105",
    number = "3",
    pages = "035022",
    year = "2022"
}

@article{Cerdeno:2023kqo,
    author = "Cerde{\~n}o, David G. and Cerme{\~n}o, Marina and Farzan, Yasaman",
    title = "{Constraints from the duration of supernova neutrino burst on on-shell light gauge boson production by neutrinos}",
    eprint = "2301.00661",
    archivePrefix = "arXiv",
    primaryClass = "hep-ph",
    reportNumber = "IFT-UAM/CSIC-22-130; FTUAM-22-2",
    doi = "10.1103/PhysRevD.107.123012",
    journal = "Phys. Rev. D",
    volume = "107",
    number = "12",
    pages = "123012",
    year = "2023"
}

@article{Akita:2023iwq,
    author = "Akita, Kensuke and Im, Sang Hui and Masud, Mehedi and Yun, Seokhoon",
    title = "{Limits on heavy neutral leptons, Z$^{′}$ bosons and majorons from high-energy supernova neutrinos}",
    eprint = "2312.13627",
    archivePrefix = "arXiv",
    primaryClass = "hep-ph",
    reportNumber = "CTPU-PTC-23-55",
    doi = "10.1007/JHEP07(2024)057",
    journal = "JHEP",
    volume = "07",
    pages = "057",
    year = "2024"
}

@article{Lai:2024mse,
    author = "Lai, Kwang-Chang and Leung, Chun Sing Jason and Lin, Guey-Lin",
    title = "{SN1987A constraints to BSM models with extra neutral bosons near the trapping regime: U(1)L{\ensuremath{\mu}}-L{\ensuremath{\tau}} model as an illustrative example}",
    eprint = "2401.16023",
    archivePrefix = "arXiv",
    primaryClass = "hep-ph",
    doi = "10.1103/PhysRevD.110.103023",
    journal = "Phys. Rev. D",
    volume = "110",
    number = "10",
    pages = "103023",
    year = "2024"
}

@article{Brune:2018sab,
    author = {Brune, Tim and P{\"a}s, Heinrich},
    title = "{Massive Majorons and constraints on the Majoron-neutrino coupling}",
    eprint = "1808.08158",
    archivePrefix = "arXiv",
    primaryClass = "hep-ph",
    reportNumber = "DO-TH 18/23",
    doi = "10.1103/PhysRevD.99.096005",
    journal = "Phys. Rev. D",
    volume = "99",
    number = "9",
    pages = "096005",
    year = "2019"
}

@article{Heurtier:2016otg,
    author = "Heurtier, Lucien and Zhang, Yongchao",
    title = "{Supernova Constraints on Massive (Pseudo)Scalar Coupling to Neutrinos}",
    eprint = "1609.05882",
    archivePrefix = "arXiv",
    primaryClass = "hep-ph",
    reportNumber = "ULB-TH-16-16",
    doi = "10.1088/1475-7516/2017/02/042",
    journal = "JCAP",
    volume = "02",
    pages = "042",
    year = "2017"
}

@article{Chen:2022kal,
    author = "Chen, Yu-Ming and Sen, Manibrata and Tangarife, Walter and Tuckler, Douglas and Zhang, Yue",
    title = "{Core-collapse supernova constraint on the origin of sterile neutrino dark matter via neutrino self-interactions}",
    eprint = "2207.14300",
    archivePrefix = "arXiv",
    primaryClass = "hep-ph",
    doi = "10.1088/1475-7516/2022/11/014",
    journal = "JCAP",
    volume = "11",
    pages = "014",
    year = "2022"
}

@article{Diamond:2023scc,
    author = "Diamond, Melissa and Fiorillo, Damiano F. G. and Marques-Tavares, Gustavo and Vitagliano, Edoardo",
    title = "{Axion-sourced fireballs from supernovae}",
    eprint = "2303.11395",
    archivePrefix = "arXiv",
    primaryClass = "hep-ph",
    doi = "10.1103/PhysRevD.107.103029",
    journal = "Phys. Rev. D",
    volume = "107",
    number = "10",
    pages = "103029",
    year = "2023",
    note = "[Erratum: Phys.Rev.D 108, 049902 (2023)]"
}

@article{Antel:2023hkf,
    author = "Antel, C. and others",
    title = "{Feebly-interacting particles: FIPs 2022 Workshop Report}",
    eprint = "2305.01715",
    archivePrefix = "arXiv",
    primaryClass = "hep-ph",
    reportNumber = "CERN-TH-2023-061, DESY-23-050, FERMILAB-PUB-23-149-PPD, INFN-23-14-LNF, JLAB-PHY-23-3789, LA-UR-23-21432, MITP-23-015",
    doi = "10.1140/epjc/s10052-023-12168-5",
    journal = "Eur. Phys. J. C",
    volume = "83",
    number = "12",
    pages = "1122",
    year = "2023"
}

@article{Cappiello:2025tws,
    author = "Cappiello, Christopher V. and Dev, P. S. Bhupal and Patwardhan, Amol V.",
    title = "{New Supernova Constraints on Neutrinophilic Dark Sector}",
    eprint = "2503.09691",
    archivePrefix = "arXiv",
    primaryClass = "hep-ph",
    month = "3",
    year = "2025"
}

@article{Fiorillo:2022cdq,
    author = "Fiorillo, Damiano F. G. and Raffelt, Georg G. and Vitagliano, Edoardo",
    title = "{Strong Supernova 1987A Constraints on Bosons Decaying to Neutrinos}",
    eprint = "2209.11773",
    archivePrefix = "arXiv",
    primaryClass = "hep-ph",
    doi = "10.1103/PhysRevLett.131.021001",
    journal = "Phys. Rev. Lett.",
    volume = "131",
    number = "2",
    pages = "021001",
    year = "2023"
}

@article{Caputo:2022mah,
    author = "Caputo, Andrea and Janka, Hans-Thomas and Raffelt, Georg and Vitagliano, Edoardo",
    title = "{Low-Energy Supernovae Severely Constrain Radiative Particle Decays}",
    eprint = "2201.09890",
    archivePrefix = "arXiv",
    primaryClass = "astro-ph.HE",
    doi = "10.1103/PhysRevLett.128.221103",
    journal = "Phys. Rev. Lett.",
    volume = "128",
    number = "22",
    pages = "221103",
    year = "2022"
}

@article{Caputo:2022rca,
    author = "Caputo, Andrea and Raffelt, Georg and Vitagliano, Edoardo",
    title = "{Radiative transfer in stars by feebly interacting bosons}",
    eprint = "2204.11862",
    archivePrefix = "arXiv",
    primaryClass = "astro-ph.SR",
    doi = "10.1088/1475-7516/2022/08/045",
    journal = "JCAP",
    volume = "08",
    number = "08",
    pages = "045",
    year = "2022"
}

@article{Fiorillo:2024upk,
    author = "Fiorillo, Damiano F. G. and Vitagliano, Edoardo",
    title = "{Self-Interacting Dark Sectors in Supernovae Can Behave as a Relativistic Fluid}",
    eprint = "2404.07714",
    archivePrefix = "arXiv",
    primaryClass = "hep-ph",
    doi = "10.1103/PhysRevLett.133.251004",
    journal = "Phys. Rev. Lett.",
    volume = "133",
    number = "25",
    pages = "251004",
    year = "2024"
}

@article{Fabbrichesi:2020wbt,
    author = "Fabbrichesi, Marco and Gabrielli, Emidio and Lanfranchi, Gaia",
    title = "{The Dark Photon}",
    eprint = "2005.01515",
    archivePrefix = "arXiv",
    primaryClass = "hep-ph",
    doi = "10.1007/978-3-030-62519-1",
    month = "5",
    year = "2020"
}

@article{Holdom:1985ag,
    author = "Holdom, Bob",
    title = "{Two U(1)'s and Epsilon Charge Shifts}",
    reportNumber = "UTPT-85-30",
    doi = "10.1016/0370-2693(86)91377-8",
    journal = "Phys. Lett. B",
    volume = "166",
    pages = "196--198",
    year = "1986"
}

@article{An:2013yfc,
    author = "An, Haipeng and Pospelov, Maxim and Pradler, Josef",
    title = "{New stellar constraints on dark photons}",
    eprint = "1302.3884",
    archivePrefix = "arXiv",
    primaryClass = "hep-ph",
    reportNumber = "PI-PARTPHYS-318",
    doi = "10.1016/j.physletb.2013.07.008",
    journal = "Phys. Lett. B",
    volume = "725",
    pages = "190--195",
    year = "2013"
}

@article{Caputo:2025aac,
    author = "Caputo, Andrea and Janka, Hans-Thomas and Raffelt, Georg and Yun, Seokhoon",
    title = "{Cooling the Shock: New Supernova Constraints on Dark Photons}",
    eprint = "2502.01731",
    archivePrefix = "arXiv",
    primaryClass = "hep-ph",
    doi = "10.1103/PhysRevLett.134.151002",
    journal = "Phys. Rev. Lett.",
    volume = "134",
    number = "15",
    pages = "151002",
    year = "2025"
}

@article{Zhang:2014wra,
    author = "Zhang, Yue",
    title = "{Supernova Cooling in a Dark Matter Smog}",
    eprint = "1404.7172",
    archivePrefix = "arXiv",
    primaryClass = "hep-ph",
    reportNumber = "CALT-68-2888",
    doi = "10.1088/1475-7516/2014/11/042",
    journal = "JCAP",
    volume = "11",
    pages = "042",
    year = "2014"
}

@article{Janka:2006fh,
    author = "Janka, Hans-Thomas and Langanke, K. and Marek, A. and Martinez-Pinedo, G. and Mueller, B.",
    title = "{Theory of Core-Collapse Supernovae}",
    eprint = "astro-ph/0612072",
    archivePrefix = "arXiv",
    doi = "10.1016/j.physrep.2007.02.002",
    journal = "Phys. Rept.",
    volume = "442",
    pages = "38--74",
    year = "2007"
}

@article{BaBar:2017tiz,
    author = "Lees, J. P. and others",
    collaboration = "BaBar",
    title = "{Search for Invisible Decays of a Dark Photon Produced in ${e}^{+}{e}^{-}$ Collisions at BaBar}",
    eprint = "1702.03327",
    archivePrefix = "arXiv",
    primaryClass = "hep-ex",
    reportNumber = "BABAR-PUB-17-001, SLAC-PUB-16923",
    doi = "10.1103/PhysRevLett.119.131804",
    journal = "Phys. Rev. Lett.",
    volume = "119",
    number = "13",
    pages = "131804",
    year = "2017"
}

@article{BESIII:2017fwv,
    author = "Ablikim, M. and others",
    collaboration = "BESIII",
    title = "{Dark Photon Search in the Mass Range Between 1.5 and 3.4 GeV/$c^2$}",
    eprint = "1705.04265",
    archivePrefix = "arXiv",
    primaryClass = "hep-ex",
    doi = "10.1016/j.physletb.2017.09.067",
    journal = "Phys. Lett. B",
    volume = "774",
    pages = "252--257",
    year = "2017"
}

@article{LHCb:2017trq,
    author = "Aaij, Roel and others",
    collaboration = "LHCb",
    title = "{Search for Dark Photons Produced in 13 TeV $pp$ Collisions}",
    eprint = "1710.02867",
    archivePrefix = "arXiv",
    primaryClass = "hep-ex",
    reportNumber = "LHCB-PAPER-2017-038, CERN-EP-2017-248",
    doi = "10.1103/PhysRevLett.120.061801",
    journal = "Phys. Rev. Lett.",
    volume = "120",
    number = "6",
    pages = "061801",
    year = "2018"
}

@article{NA64:2019auh,
    author = "Banerjee, D. and others",
    collaboration = "NA64",
    title = "{Improved limits on a hypothetical X(16.7) boson and a dark photon decaying into $e^+e^-$ pairs}",
    eprint = "1912.11389",
    archivePrefix = "arXiv",
    primaryClass = "hep-ex",
    reportNumber = "CERN-EP-2019-284",
    doi = "10.1103/PhysRevD.101.071101",
    journal = "Phys. Rev. D",
    volume = "101",
    number = "7",
    pages = "071101",
    year = "2020"
}

@article{Bauer:2018onh,
    author = "Bauer, Martin and Foldenauer, Patrick and Jaeckel, Joerg",
    title = "{Hunting All the Hidden Photons}",
    eprint = "1803.05466",
    archivePrefix = "arXiv",
    primaryClass = "hep-ph",
    doi = "10.1007/JHEP07(2018)094",
    journal = "JHEP",
    volume = "07",
    pages = "094",
    year = "2018"
}

@article{Pospelov:2008zw,
    author = "Pospelov, Maxim",
    title = "{Secluded U(1) below the weak scale}",
    eprint = "0811.1030",
    archivePrefix = "arXiv",
    primaryClass = "hep-ph",
    doi = "10.1103/PhysRevD.80.095002",
    journal = "Phys. Rev. D",
    volume = "80",
    pages = "095002",
    year = "2009"
}

@article{Jaeckel:2010ni,
    author = "Jaeckel, Joerg and Ringwald, Andreas",
    title = "{The Low-Energy Frontier of Particle Physics}",
    eprint = "1002.0329",
    archivePrefix = "arXiv",
    primaryClass = "hep-ph",
    reportNumber = "CPT-10-18, DESY-10-016, IPPP-10-09",
    doi = "10.1146/annurev.nucl.012809.104433",
    journal = "Ann. Rev. Nucl. Part. Sci.",
    volume = "60",
    pages = "405--437",
    year = "2010"
}

@article{Essig:2015cda,
    author = "Essig, Rouven and Fernandez-Serra, Marivi and Mardon, Jeremy and Soto, Adrian and Volansky, Tomer and Yu, Tien-Tien",
    title = "{Direct Detection of sub-GeV Dark Matter with Semiconductor Targets}",
    eprint = "1509.01598",
    archivePrefix = "arXiv",
    primaryClass = "hep-ph",
    doi = "10.1007/JHEP05(2016)046",
    journal = "JHEP",
    volume = "05",
    pages = "046",
    year = "2016"
}

@article{Zentner:2009is,
    author = "Zentner, Andrew R.",
    title = "{High-Energy Neutrinos From Dark Matter Particle Self-Capture Within the Sun}",
    eprint = "0907.3448",
    archivePrefix = "arXiv",
    primaryClass = "astro-ph.HE",
    doi = "10.1103/PhysRevD.80.063501",
    journal = "Phys. Rev. D",
    volume = "80",
    pages = "063501",
    year = "2009"
}

@article{Chen:2014hha,
    author = "Chen, Chian-Shu and Lee, Fei-Fan and Lin, Guey-Lin and Lin, Yen-Hsun",
    editor = "Aguilar-Benitez, M and Fuster, J and Marti-Garcia, S and Santamaria, A",
    title = "{The dark matter self-interaction and its impact on the critical mass for dark matter evaporations inside the sun}",
    eprint = "1412.6739",
    archivePrefix = "arXiv",
    primaryClass = "hep-ph",
    doi = "10.1016/j.nuclphysbps.2015.09.049",
    journal = "Nucl. Part. Phys. Proc.",
    volume = "273-275",
    pages = "347--352",
    year = "2016"
}

\end{document}